\documentclass[manuscript]{acmart}
\usepackage{amsmath}
\usepackage{mathtools}
\usepackage{graphicx}
\usepackage{subfiles}
\usepackage{subcaption}
\usepackage{xcolor}
\usepackage{float}
\usepackage[skip=4pt]{caption}
\usepackage{appendix}
\usepackage{rotating}
\usepackage{tabularx} 
\usepackage{caption}
\usepackage{todonotes}

\newcommand{\appref}[1]{Appendix~\ref{#1}}

\AtBeginDocument{%
  }

\setcopyright{acmlicensed}
\copyrightyear{2024}
\acmYear{2024}
\acmDOI{XXXXXXX.XXXXXXX}

\acmConference[Conference acronym 'XX]{Make sure to enter the correct conference title from your rights confirmation email}{June 03--05,
  2018}{Woodstock, NY}

\acmISBN{978-1-4503-XXXX-X/18/06}
\graphicspath{{Figures/}}

\begin{document}

\title{ParaTutor: Coordinating Parent-Child Math Tutoring through Role-Separated LLM Scaffolding}

% \author{Anonymous author(s)}
\author{LUO Lan}
\email{lluo476@connect.hkust-gz.edu.cn}
\affiliation{%
  \institution{Hong Kong University of Science and Technology(Guangzhou)}
  \city{Guangzhou}
  \country{China}
}

\author{WANG Anqi}
\email{awangan@connect.ust.hk}
\affiliation{%
  \institution{Hong Kong University of Science and Technology}
  \city{Hong Kong SAR}
  \country{Hong Kong SAR}
}

\author{ZHOU Muzhi}
\email{mzzhou@ust.hk}
\affiliation{%
  \institution{Hong Kong University of Science and Technology(Guangzhou)}
  \city{Guangzhou}
  \country{China}
}

\author{ZHU Junhua}
\email{junhuazhu@hkust-gz.edu.cn}
\affiliation{%
  \institution{Hong Kong University of Science and Technology(Guangzhou)}
  \city{Guangzhou}
  \country{China}
}

\author{CAI Jie}
\email{jie.cai1@outlook.com}
\affiliation{%
  \institution{Tsinghua University}
  \city{Beijing}
  \country{China}
}

\author{YU Ao}
\email{aoyu@hkust-gz.edu.cn}
\affiliation{%
  \institution{Hong Kong University of Science and Technology(Guangzhou)}
  \city{Guangzhou}
  \country{China}
}

\author{PAN Hui}
\email{panhui@ust.hk}
\affiliation{%
  \institution{Hong Kong University of Science and Technology(Guangzhou)}
  \city{Guangzhou}
  \country{China}
}

\begin{abstract}
Parent-child tutoring is a collaborative learning setting with asymmetric roles. Parents guide children’s problem solving, while children are expected to remain actively engaged in understanding and reasoning. However, most LLM-based learning systems are designed for single users or relatively symmetric collaboration, leaving parent--child tutoring with distinct instructional roles underexplored. Through a formative study, we found that parent-child math tutoring was often disrupted by cognitive misalignment, emotional escalation, and method mismatch. To address these challenges, we present \textbf{ParaTutor}, a multi-agent LLM-based scaffolding system for home math word-problem tutoring. ParaTutor distributes support across user roles by providing parents with strategy, language, repair, and phase scaffolds, while providing children with visual grounding for problem interpretation. We evaluated ParaTutor with 23 parent--child dyads (children aged 10--12) across four tutoring conditions that varied how LLM assistance was delivered. Results show that generic LLM assistance often provided useful explanations but did not consistently support parent-led tutoring or children’s active reasoning. In contrast, ParaTutor helped redistribute tutoring work across parents and children, increased children’s engagement with word problems, supported shared understanding through visual grounding, and helped parents translate LLM-generated methods into child-facing tutoring moves. These findings suggest that in family learning, the value of LLM support depends not only on model capability, but also on how support is coordinated across users with different roles. Our work contributes design implications for LLM systems that support role-sensitive scaffolding in parent-child learning.
\end{abstract}

\begin{teaserfigure}
    \centering
    \includegraphics[width=0.95\textwidth]{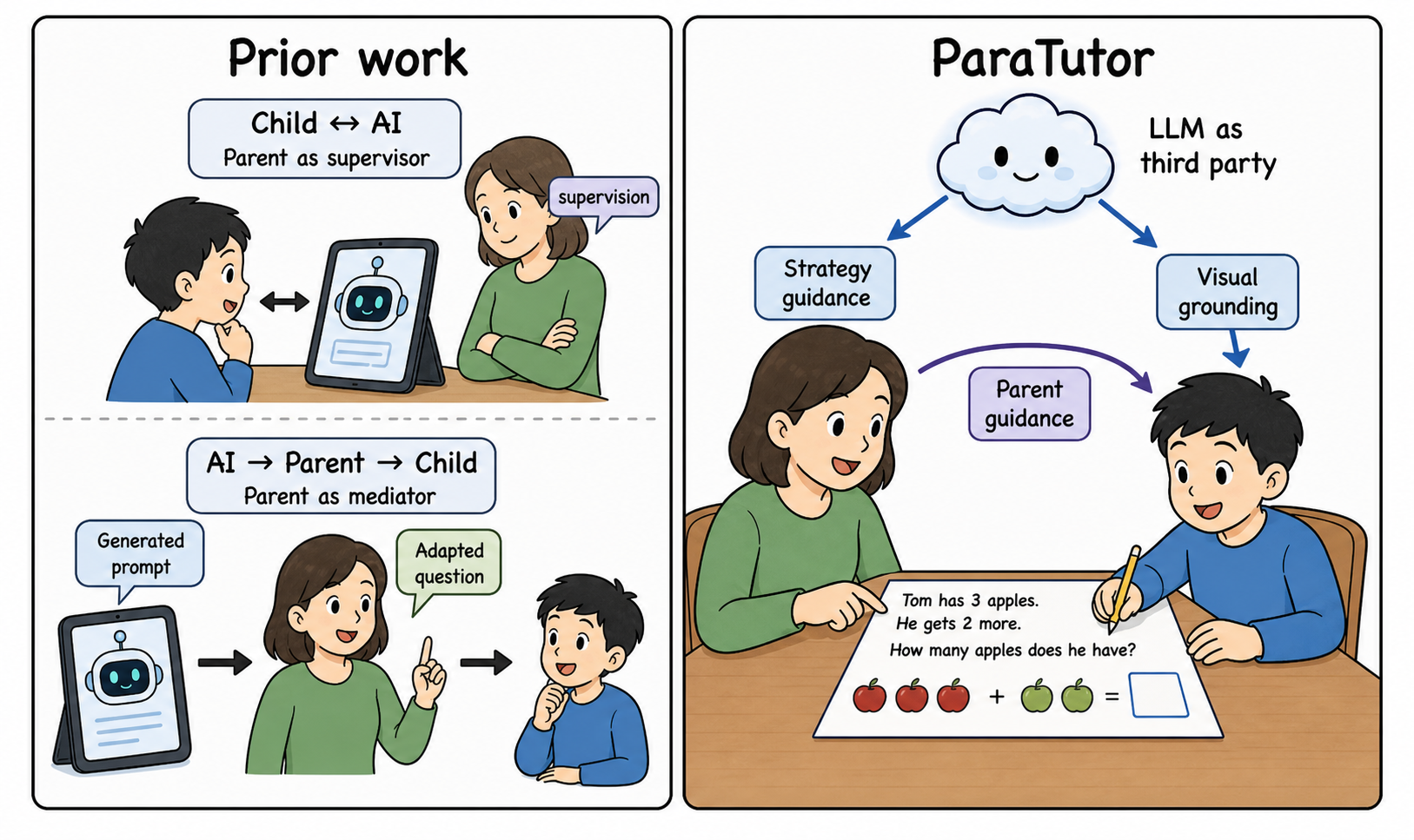}
    \Description{A two panel illustration comparing prior AI mediated parent child learning with ParaTutor. The prior work panel shows AI providing prompts to a parent, who mediates them to the child. The ParaTutor panel shows the LLM as a third party that provides strategy guidance to the parent and visual grounding to the child during math tutoring.}
    \caption{ParaTutor positions the LLM as a third party in parent guided math word problem tutoring. The system provides parents with strategy guidance and children with problem solving support while preserving parents’ guiding role in the tutoring interaction.}
    \label{fig:conceptual-overview}
\end{teaserfigure}

\begin{CCSXML}
<ccs2012>
   <concept>
       <concept_id>10010405.10010489.10010490</concept_id>
       <concept_desc>Applied computing~Computer-assisted instruction</concept_desc>
       <concept_significance>500</concept_significance>
       </concept>
   <concept>
       <concept_id>10003120.10003121.10003129</concept_id>
       <concept_desc>Human-centered computing~Interactive systems and tools</concept_desc>
       <concept_significance>500</concept_significance>
       </concept>
   <concept>
       <concept_id>10003120.10003121.10011748</concept_id>
       <concept_desc>Human-centered computing~Empirical studies in HCI</concept_desc>
       <concept_significance>300</concept_significance>
       </concept>
 </ccs2012>
\end{CCSXML}

\ccsdesc[500]{Applied computing~Computer-assisted instruction}
\ccsdesc[500]{Human-centered computing~Interactive systems and tools}
\ccsdesc[300]{Human-centered computing~Empirical studies in HCI}

\keywords{Human-centered Computing, Interactive Systems and Tools, Large Language Models, Parental Involvement, Mathematical Word Problem}

\maketitle

\section{Introduction}
In China, parental involvement in children’s homework is a routine part of family educational life.  More broadly, parental tutoring extends academic experiences beyond the classroom and provides essential support for cognitive development and achievement with targeted feedback \cite{jaiswal2017review}. However, parental tutoring has often been treated as a general construct, with less attention to how parental assistance differs across subject domains~\cite{gonzalez2005examining, boonk2018review, djurivsic2017parental}. This distinction is important in mathematics, where word problems are among the most cognitively demanding tasks because they require children to integrate linguistic comprehension with mathematical reasoning~\cite{zahrah2020contextual}. For upper elementary students in China, solving word problems is a central part of mathematics learning because it supports conceptual understanding, analytical and creative thinking, and the application of mathematics to real life situations~\cite{verschaffel2000making, frankenstein2009developing, bodovski2007mathematics}. Parental tutoring in this area therefore requires parents not only to understand problem structures, but also to move beyond their own familiar reasoning habits and explain mathematical relations in ways that children can follow. Many parents consequently find math word problems one of the most challenging tasks to tutor. Gaps between adult perspectives and children’s understanding can create tension during tutoring and may further diminish children’s interest and motivation in learning math~\cite{purnomo2022mother, mutangira2024early}.

Large language models (LLMs) offer new opportunities to support math tutoring at home. By delivering explanations and feedback in natural language, LLMs can make complex concepts more accessible and help learners engage with difficult problems \cite{ho2023designing, zhang2022storybuddy}.  
However, supporting math tutoring at home requires more than providing explanations to an individual learner, because parents and children participate in the tutoring process with different roles.
Prior work on LLM supported family learning has often positioned LLMs as direct partners for children, sometimes with parental supervision \cite{chen2025characterizing, he2025storypal, wang2025charactercritique, yang2026autiverse,xie2026understanding} or provide parents with resources to mediate children’s learning ~\cite{dietz2024contextq, ho2025set, dangol2025want, nawshin2026well}. 
More recent work such as AACessTalk has begun to support both parents and children within the same interaction, providing guide messages for parents and vocabulary card recommendations for children to facilitate communication between minimally verbal autistic children and parents~\cite{choi2025aacesstalk}. However, this work focuses on communication access rather than instructional scaffolding. Less is known about how LLM support can be differentiated across parents and children when both participate in the same homework tutoring task, where parents need tutoring strategies and children need support for understanding and reasoning.
Therefore, rather than designing an LLM tutor for children to use independently, we investigate how an LLM can act as a third party in parent guided math word problem tutoring by supporting parents’ guiding and children’s problem solving in different ways.

To address this gap, we propose an LLM system that supports parent child math word problem tutoring by scaffolding parents’ guiding and children’s problem solving in different ways. Rather than replacing parental involvement with a child facing tutor, our goal is to preserve the pedagogical value of parental tutoring while helping parents organize, adapt, and deliver tutoring strategies during interaction. We draw on scaffolding as a design lens to examine how support can be distributed across the LLM system, parents, and children in home tutoring. To reach this goal, we ask three questions:
\begin{itemize}
\item How do parents scaffold children’s problem solving, and what challenges emerge during everyday math homework tutoring in China?
\item How can role-separated and phase-gated LLM scaffolding be designed to support parent-led tutoring while preserving children’s active reasoning?
\item How does \textit{ParaTutor} affect parent-child tutoring interactions, including children’s engagement, shared understanding, and parents’ use of LLM scaffolds?
\end{itemize}

We conducted a formative study in the Chinese home tutoring context to understand common math word problem solving procedures, challenges in parental tutoring, and tutoring strategies adopted by parents. Based on these findings, we developed \textit{ParaTutor}, an LLM based system that uses multiple agents to provide role separated scaffolding for parent child tutoring. ParaTutor supports interactive problem solving for a given parent child dyad by offering parents strategy level guidance and children problem level support tailored to the tutoring context.
We then conducted a user study with 23 parent child dyads to evaluate the prototype. The study examined how ParaTutor shaped tutoring interactions across different parental tutoring strategies and how parents and children experienced the system during math word problem tutoring. Our findings show how role-separated and phase-gated LLM scaffolding can redistribute tutoring work across parents and children, support children’s engagement and shared understanding, and help parents translate LLM-generated methods into child-facing tutoring moves.
The study protocol of this project was approved by the university’s Institutional Review Board (IRB), and informed consent was obtained from all participating parents.

\section{RELATED WORK}
\subsection{Parental tutoring as Collaborative Family Work in China}
Parental tutoring refers to home-based instructional support in which parents assist children with academic tasks such as homework \cite{taylor2016teacher}. In this setting, parents do more than check answers. They help structure the learning process by setting the pace, explaining problem statements, prompting children to reason, correcting mistakes, and deciding when to step in or step back. Although parents’ familiarity with their children can enable personalized support \cite{nickow2020impressive}, tutoring can also become tense when parents and children differ in how they understand a problem or what method should be used. Parents are not trained educators, and they may lack up-to-date subject knowledge, pedagogical strategies, or representational tools for explaining abstract concepts \cite{nickow2020transformative, gao2024parent}. As a result, parent-child tutoring is often both cognitive and interactional work.

These challenges become especially salient in math word problems. Unlike routine exercises, word problems require children to interpret narrative descriptions, identify relevant quantities, map relationships to mathematical representations, and maintain intermediate reasoning steps \cite{kintsch1985understanding, verschaffel2020word, swanson2008growth, zheng2011working}. Problem solving also unfolds across phases, such as understanding the situation, planning a solution, executing calculations, and reflecting on results \cite{polya2014solve, jin2024using}. Each phase requires different forms of support. Parents may need to help children understand what the problem is asking before calculation, guide reasoning without taking over, and later help children summarize the method. This makes math word problems a useful context for studying how parental guidance is coordinated during home tutoring.

In China, these tutoring difficulties are shaped by broader expectations around parental responsibility for children’s academic success. Research on Chinese parenting has emphasized the cultural notion of ``training,'' where parents are expected to actively guide, correct, and supervise children’s learning rather than provide only optional assistance \cite{chao1994beyond, chao1996chinese}. Recent HCI work on Chinese homework interactions further shows that homework tutoring often involves emotional strain, knowledge conflicts, and parent-child tension rather than only instructional support \cite{gao2025homeworkwars}. In this context, when parents lack the knowledge, strategy, or language needed to guide children, tutoring breakdowns can affect both learning progress and the parent’s role as a capable guide. This motivates our focus on systems that support not only children’s problem solving, but also parents’ ability to coordinate tutoring.

\subsection{LLM-Mediated Support in Family Learning}
LLMs have increasingly been introduced into educational settings as conversational tutors and learning companions \cite{chen2025cograder,cheng2025oak}. Their ability to generate step-by-step explanations, respond to follow-up questions, and adapt content dynamically enables more flexible support than earlier rule-based systems \cite{Lyu_2024,nye2014autotutor,venugopalan2025combining}. In mathematics education, LLM-based systems show promise in interpreting problem statements, producing reasoning steps, and offering structured guidance \cite{li2020graph,zhang2020graph,jie2022learning}. These capabilities make AI-supported tutoring increasingly feasible for everyday learning contexts. 

Recent HCI research has explored multiple ways of introducing LLM into family learning interactions. One common configuration positions LLM as a direct partner for children, with parents supervising or evaluating children’s use of the system. For example, StoryPal supports young children’s dialogic reading with an LLM, while CharacterCritique uses multi-agent interaction to support children’s critical thinking during story reading~\cite{he2025storypal, wang2025charactercritique}. Another configuration provides parents with LLM-generated resources that they can select, adapt, or mediate for children. ContextQ, for instance, generates dialogic questions for parents to use during co-reading~\cite{dietz2024contextq}, while BrickSmart provides parents with guidance for supporting children’s spatial language learning during block play~\cite{liu2025bricksmart}. These systems show that LLM can enter family learning either by interacting directly with children or by supporting parents as mediators and facilitators.

Less is known about how LLM support can be differentiated across parents and children within the same tutoring interaction. Some systems have begun to move in this direction. For example, AACessTalk provides parents with guide messages and children with vocabulary card recommendations to support communication between minimally verbal autistic children and parents~\cite{choi2025aacesstalk}. Yet this work focuses on communication access rather than instructional scaffolding in homework tutoring. Math homework tutoring presents a different coordination problem because parents guide the process while children need to remain actively engaged in understanding and solving problems. Our work examines this configuration by positioning the LLM as a third party in parent-child math word problem tutoring, providing strategy guidance to parents and visual grounding to children while keeping parents in the guiding role.

\subsection{Scaffolding as Role Coordination in Asymmetric Learning Interaction}

Scaffolding refers to guided support provided by a more experienced partner to help a learner accomplish tasks that would otherwise be difficult to complete independently \cite{vygotsky1978mind,rogoff1990apprenticeship}. Rather than simply giving answers, scaffolding involves regulating when and how support is introduced, adapting explanations to the learner’s current understanding, and gradually withdrawing assistance as competence develops \cite{van2010scaffolding}. In this sense, scaffolding functions not only as a pedagogical strategy but also as an interactional mechanism that shapes turn-taking, pacing, attention, and the distribution of responsibility during problem solving.

Prior HCI research has increasingly operationalized scaffolding in interactive systems. Existing systems use structured prompts \cite{lee2022promptiverse,liu2024selenite}, decomposition \cite{kim2018agile,bhattacharjee2024understanding} and annotation \cite{lee2022promptiverse} to support users in tasks such as writing \cite{dhillon2024shaping,hui2023lettersmith}, programming \cite{ma2025dbox}, literature review \cite{palani2023relatedly}, and healthcare communication \cite{hu2024designing}. These systems show that scaffolding can help users externalize intermediate structure, stay oriented in complex tasks, and receive context-sensitive support without immediately giving final answers. However, most of this work focuses on single-user learning or human-AI collaboration. By contrast, parent-child tutoring is an asymmetric and relational setting, where scaffolding must support parents in guiding children’s reasoning across different tutoring phases without taking over the problem-solving process. In response, we examine how scaffolding can be organized across tutoring phases in parent-child learning, and present a system that supports parents in providing timely, adaptive guidance while keeping children actively engaged in problem solving.

\section{FORMATIVE STUDY}
\label{sec:FormativeStudy}
To address RQ1, we conducted a formative study to understand how parent and child roles are organized during math word problem tutoring, and what challenges disrupt this role structure.

\subsection{Participants and Procedure}
We recruited eleven parents (P1–P11) and two experienced primary school teachers (T1, T2). Participants were recruited through online parent communities, including \textit{Xiaohongshu}, a popular social media platform in China, as well as local after-school institutions. All participating parents were primary caregivers of children in Grades 4–6 (ages 10–12) and regularly involved in home tutoring. The two teachers had 26 and 38 years of teaching experience in upper-elementary mathematics. Table \ref{tab:formativestudy-participants} presents background information about the parent participants and their children.

We conducted semi-structured interviews with teachers and parents to examine instructional practices in school and home settings. Each interview lasted for 20 to 30 minutes. Interview questions included how tutoring was structured across problem-solving phases, how problem-solving responsibilities were distributed between parents and children, and where difficulties or breakdowns occurred. We also explored the strategies parents used to guide learning and how interaction unfolded in practice, including how turns were coordinated, how intervention timing was decided, and how artifacts (e.g., drawings, phone searches, scratch paper) were used to maintain shared understanding. All interviews were audio-recorded with consent and anonymized for analysis.

\begin{table}[!h]
    \renewcommand{\arraystretch}{1.2}
    \caption{Basic Information of Parents and Children in the Formative Study}
    \label{tab:formativestudy-participants}
    \centering
    \resizebox{\columnwidth}{!}{
    \begin{tabular}{c c c c c c}
        \toprule
        ID  & Child's Gender & Grade & Usual Math Grades & Tutoring Difficulty & Recruitment Mode \\
        \midrule
        P1  & M & 6th & Excellent & Easy                & Offline \\
        P2  & M & 6th & Excellent & Relatively Easy     & Online \\
        P3  & M & 6th & Excellent & Relatively Easy     & Online \\
        P4  & F & 6th & Excellent & Easy                & Offline \\
        P5  & F & 5th & Excellent & Relatively Difficult& Online \\
        P6  & M & 4th & Excellent & Easy                & Online \\
        P7  & F & 4th & Good      & Easy                & Offline \\
        P8  & M & 4th & Medium    & Relatively Difficult& Offline \\
        P9  & M & 5th & Excellent & Easy                & Offline \\
        P10 & F & 4th & Excellent & Easy                & Online \\
        P11 & M & 4th & Excellent & Relatively Easy     & Online \\
        \bottomrule
    \end{tabular}
    }
\end{table}

\subsection{Data Analysis}
We conducted a thematic analysis. Two researchers independently reviewed the transcripts and generated initial codes capturing phenomena such as role responsibilities, intervention timing, phase transitions, artifact use, and strategy mismatch. Through iterative comparison and discussion, the researchers developed a shared codebook with definitions and representative excerpts. Discrepancies were resolved in team meetings to ensure analytic consistency. Our analysis aimed to catalog tutoring strategies and identify recurring breakdown patterns and coordination mechanisms in the parental tutoring process. The notion of instructional asymmetry emerged through iterative coding of role distribution and phase transitions.

\subsection{Findings}
Across interviews, we identified recurring patterns in how parents and children maintain distinct instructional roles during math tutoring, as well as situations where these roles become blurred or disrupted. 
Overall, effective tutoring relied on parents guiding the learning process while children remained responsible for reasoning and solution generation.

\subsubsection{Distinct Roles Are Maintained Through Staged Tutoring}
\label{sec:WordProblemSolvingSteps}
Across both teacher and parent interviews, word problem solving was described as a process that embodies three recurrent phases, namely \textbf{understanding}, \textbf{calculation}, and \textbf{summarization}, and aligns with classical problem solving models~\cite{polya2014solve}. Each phase establishes a temporary focus of attention and a shared expectation of what counts as progress, thereby regulating when reasoning advances and how instructional authority and responsibility are distributed in interaction.

The \textbf{understanding} phase begins with the parent and child working toward shared interpretive alignment. As T1 stressed, students are required to "read the question three times without adding or omitting any words" before attempting computation. Parents similarly reported authorizing phase transitions, deciding whether comprehension was adequate to proceed (P5, P10, P11). Some described asking children to restate the problem in their own words before any calculation (P4, P7). \textbf{Calculation} typically proceeds after shared \textbf{understanding} is established. When errors occur, T2 reported that students are instructed to "return to the question instead of correcting numbers blindly", thus going back to the understanding phase. Parents described similar strategies (P2). The \textbf{summarization} phase extends beyond answer verification. Teachers and parents both described prompting children to articulate underlying principles and consider how the method applies to other similar problems (P3, P6). 
%As T2 noted, without structured staging, older students may "directly copy the complete answer" rather than reason step by step.

Based on accounts from teachers and parents, math word problem solving is organized around these clearly defined stages. These phase boundaries are actively enforced rather than implicitly assumed. Parents often require children to proceed through these stages in sequence, using them as an interactional script to guide instruction and regulate problem-solving behavior. This staged progression enables parents to assert pedagogical authority by directing how the problem should be approached.

% 策略带上不对称结构。
\subsubsection{Distinct Roles Are Maintained Through Scaffolding Strategies}
\label{sec:ParentsMathTutoringStrategies}
%怎么调节控制权的

We identified recurring scaffolding moves that parents use. Rather than isolated techniques, these strategies shape turn-taking, pacing, and the distribution of responsibility during problem solving.

\textbf{Understanding Phase.}
\textit{Task simplification} breaks down lengthy or complex problem statements into smaller, more manageable units. Parents noted that children often struggle with questions containing dense wording (P2, P5, P7, P9). As P2 described, "For example, use a pen to circle parts of the problem on paper, so they can gradually understand it better and eventually grasp it fully". He further explained, "Sometimes I would ask him to read out the question part by part, according to the commas." These actions make the structure of the problem more visible to the parent and child, supporting shared understanding before moving forward.

\textit{Situational guidance} refers to mapping abstract numerical relationships onto familiar real-life contexts. As P7 explained, \textit{"If a child does not know much about the specific scenarios or applications, he may not understand the meaning."} Several parents also described drawing diagrams to clarify relationships. As P3 noted, \textit{"If the question is too long, I will draw it so my child can see the relationships."} However, some parents found visualization challenging. P2 admitted, \textit{"I know drawing can really help, but I’m not good at it."} Parents sometimes replaced the original problem context with more common examples. As P8 illustrated, \textit{"For certain questions... we will use real-life examples with food… to help him understand better."}

\textbf{Calculation Phase.}
The most common strategy was \textit{feedback and correction}. This strategy can effectively prevent children from forming wrong cognitive patterns due to deviations in thinking when solving problems for the first time (P4). Especially when children have relatively weak foundations, this step-by-step explanation method is particularly important (P10). During these interactions, parents monitored not only correctness but also their child’s understanding, adjusting subsequent explanations based on the child’s responses (P7). However, this approach was not universal. Two parents preferred allowing children to complete all problems before conducting an overall review (P2, P6), believing that a holistic assessment better revealed recurring reasoning patterns (P2).

A second strategy involved is \textit{demonstration and imitation}. Four parents (P2, P3, P8, P11) described modeling a solution process when independent reasoning stalled. As P8 explained, he would first demonstrate how to solve a similar problem and then modify its values. Parents who adopted this strategy viewed imitation as a means of stabilizing reasoning patterns (P11). However, P4 expressed concern that their problem-solving approach might differ from the teachers' approach. 
% He noted that, upon reviewing the teaching requirements, he found that their current method is somewhat different from what he learned in the past.
As a regulation strategy, demonstration allows parents to take greater control temporarily and then return the task to the child when reasoning needs support.

Several parents adopted an \textit{independent-thinking strategy} to redistribute problem-solving responsibility. Rather than immediately identifying errors, they prompted children to re-examine their reasoning. As P5 stated firmly, \textit{"children must think first."} Others described asking children to articulate their solution process before offering help (P2, P6). Parents viewed this strategy as a way to strengthen independent reasoning. However, they also acknowledged that its effectiveness depends on the child’s competence. As P4 noted, even after prompting, some children struggle to progress, requiring renewed intervention. This strategy shifts more responsibility to the child while maintaining parental supervision.

\textbf{Summarization Phase.}  
The first was \textit{review and consolidation}, in which children revisited incorrect problems and articulated underlying principles. As P5 explained, he emphasized the importance of understanding "the reasons behind each type of mistake rather than just memorizing the specific errors." Rather than treating mistakes as isolated events, parents asked their children to review the question to stabilize reasoning patterns across similar problems (P3, P4, P5, P6, P7, P9, P10, P11). However, several noted that systematic review could be too time-consuming (P4, P9).

A second strategy involved \textit{practical application}. Some parents extended solutions into familiar real-life contexts, such as paper cutting (P3) or buying fruit (P8), encouraging children to apply learned methods beyond the original problem. By connecting abstract reasoning to lived experience, parents reinforced reasoning coherence.

\subsubsection{Recurring Challenges in Coordinating Parent-Child Math Tutoring}

\begin{figure}
    \centering
    \includegraphics[width=0.75\linewidth]{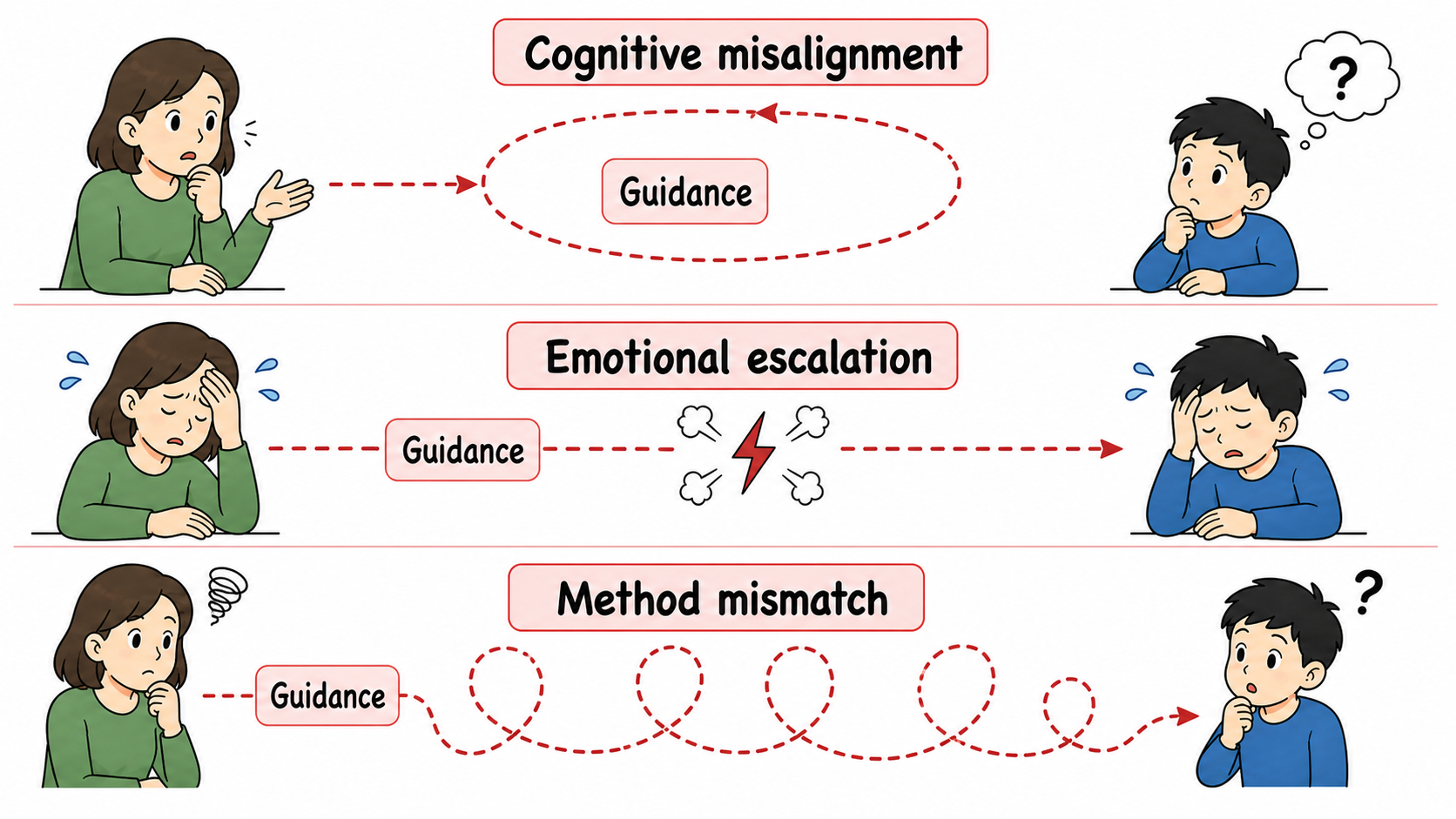}
    \caption{Recurring challenges that make parental guidance difficult to deliver.}
    \label{fig:parent_child_tutoring_challenges}
\end{figure}

Parents often encounter recurring challenges when helping their children with math. Although they are motivated to support learning, they may struggle to make problem structures visible, maintain productive communication, and align their familiar methods with the approaches currently taught at school. As shown in Figure~\ref{fig:parent_child_tutoring_challenges}, these challenges can be understood as cognitive misalignment, emotional escalation, and method mismatch. We summarize these challenges below.

\textbf{C1: Cognitive Misalignment in Externalizing Problem Structure for Children.}
Although many parents acknowledged the importance of visual representations for helping children understand math word problems, several reported difficulties in accurately expressing mathematical relationships through drawings. P2 admitted, ``I know drawing can really help, but I’m not good at it.'' Parents were often aware when their explanations were ineffective: many noticed that their children became even more confused after seeing the visual explanation. This challenge highlights a core issue in at-home tutoring: parents may recognize the value of visualization and perceive its immediate effects, yet still lack the tools to externalize problem structures in ways that are precise, interpretable, and aligned with the child’s understanding.

\textbf{C2: Emotional Escalation During Guidance.}
Parents also described communication breakdowns during tutoring (P3, P9, P10). P10 observed that children often think differently from their parents, making it difficult for children to understand the parent’s intentions or reasoning. Many parents believed that they had explained the problem clearly, yet their children remained confused and struggled to follow the explanation.

These breakdowns were often intensified by parents’ difficulty in gauging their children’s current knowledge state. Some parents, such as P3 and P10, acknowledged that their own methods might not be the most efficient, but still regarded them as more reliable than unfamiliar alternatives. Even when parents claimed to respect their children’s choices, their attitudes sometimes revealed an underlying preference for familiar approaches and skepticism toward strategies introduced by the child.

Conflicts also emerged from parents’ observations of children’s learning habits or attitudes. For example, P9 said, \textit{``Sometimes he shows no attitude, no patience, or carelessness, and starts answering questions without fully reading them.''} When children appeared impatient or careless, parents sometimes responded harshly rather than offering further guidance. Such reactions could lead to children’s resistance and further strain parent-child communication.

\textbf{C3: Method Mismatch Between Parents’ Familiar Strategies and School-Taught Approaches.}
Some parents (P2, P11) encountered knowledge gaps when tutoring their children. They had to make additional efforts to align their own understanding with the content currently taught in school. P2 explained, \textit{``I also have some concepts that are not clear, and I have to look them up because it has been so long. When I am tutoring my child, I have to look up the book myself or search on my phone.''}

In addition, 6 out of 11 parents reported that they tended to use equation-based methods that were familiar to them but beyond what their children had learned at school (P1, P2, P4, P5, P6, P7). P2 added, \textit{``My child still prefers the method taught by the teacher, which sometimes I find difficult to accept. Sometimes, I can only use my original method to explain it to him.''} This mismatch suggests that parents may have difficulty understanding or accepting newer school-taught problem-solving methods, which can in turn confuse children.

For example, P5 worried that exposing the child to multiple methods could create confusion. This concern reflects parents’ frustration with the gap between their own familiar strategies and school-based approaches, as well as their fear that introducing different methods may disrupt the child’s learning.

\section{PARATUTOR SYSTEM}
\label{MathTutor}
Instead of directly delivering answers to the child, ParaTutor is designed to support the parent’s tutoring role while keeping the child actively engaged in reasoning. The system therefore provides different forms of support to parents and children, helping parents guide tutoring without bypassing parental involvement or taking over the child’s reasoning process. This section answers RQ2.

\subsection{Design Requirements}
Our design is grounded in the formative finding that successful parent--child tutoring depends on maintaining distinct roles during interaction, in which parents guide the tutoring process and children remain engaged in reasoning. We therefore derive the following requirements directly from the recurring challenges identified in the formative study.

\textbf{R1: Provide Child-Facing Visual Grounding for Problem Understanding.}  
When parents struggle to clearly externalize mathematical relationships (C1), shared understanding can break down during early problem solving. The system should therefore generate child-facing visual representations that help make key entities and relationships visible, reducing reliance on parents' drawing skills and supporting shared problem understanding.

\textbf{R2: Support Constructive Parent-Child Communication During Tutoring.}  
When communication difficulties and emotional escalation arise between parents and children (C2), shared understanding can become difficult to maintain. The system should therefore support staged progression with shared confirmation, helping parents and children stay aligned, slowing down misunderstanding, and reducing premature correction or emotionally charged responses.

\textbf{R3: Provide Parent-Facing Tutoring Guidance That Adapts to Different Solution Methods.}  
When parents encounter knowledge gaps or differences between their familiar solution methods and school-taught approaches (C3), they may find it difficult to explain the problem or guide the child effectively. The system should therefore provide parent-facing, strategy-based guidance that helps parents support children’s reasoning and adapt to different valid ways of solving the problem.

\subsection{Scaffolding Design Across User Roles, Tutoring Phases, and Guidance Strategies}

Guided by the design requirements, ParaTutor organizes its support as a three-dimensional scaffolding design that coordinates user roles, tutoring phases, and guidance strategies during parent-child math tutoring. The role dimension differentiates the support provided to parents and children, with parents receiving guidance for conducting the tutoring interaction and children receiving visual grounding for interpreting the problem. The phase dimension structures how support changes across understanding, calculation, and summarization, so that tutoring can progress from problem interpretation to reasoning and reflection. The strategy dimension translates tutoring goals into concrete parent-facing guidance, including suggested actions, example language, and ways of using the child-facing visualization.

Together, these dimensions define ParaTutor as a scaffold for parent-led tutoring. By coordinating who receives support, when support is provided, and how parents are guided to intervene, the system supports children’s problem understanding, constructive parent-child communication, and parents’ adaptation to different solution methods. The following sections elaborate how each dimension is instantiated in the system design.

\subsubsection{Scaffolding Across User Roles}
The first dimension of ParaTutor's scaffolding concerns how system support is distributed between parents and children. Because parents often struggle to externalize mathematical relationships clearly, ParaTutor places visual grounding on the child-facing side of the interface. Children receive visual representations of key entities, quantities, and relationships in the problem, while parents receive guidance for facilitating the tutoring interaction rather than direct instruction intended for the child. This division helps establish shared problem understanding (C1), reduces reliance on parents' drawing skills, and preserves the parent’s role as the tutor.

\begin{figure*}[htp]
    \centering
    \includegraphics[width=\textwidth]{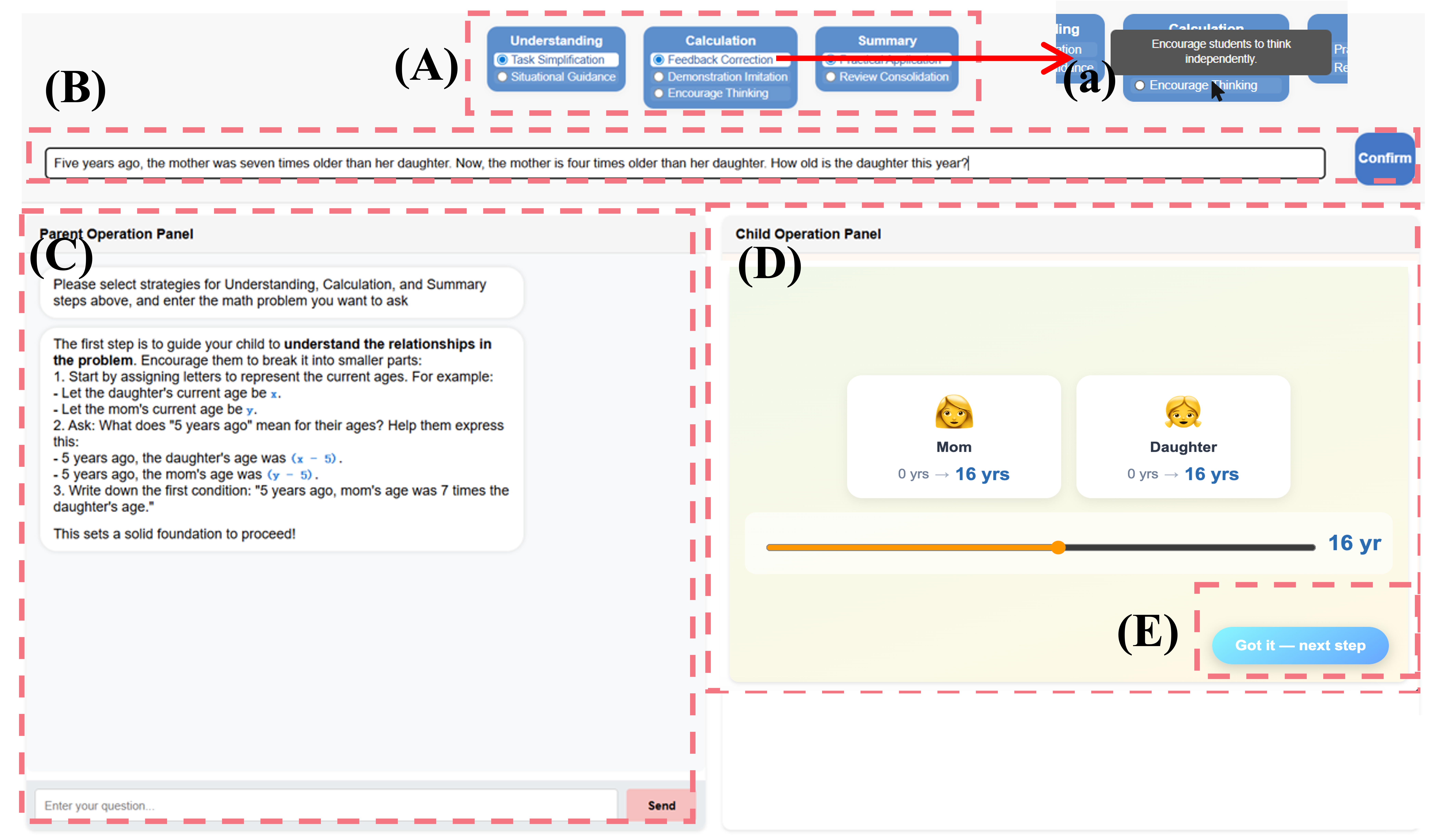}
    \caption{Dual interface of ParaTutor. The workspace is divided into a parent-facing panel (C) that provides strategy-based tutoring guidance and a child-facing panel (D) that presents visual grounding without revealing procedural solutions. Strategy selection (A) and problem input (B) configure phase-specific support. Phase progression is managed through shared confirmation (E), helping parents guide tutoring while keeping children engaged in reasoning across problem-solving stages.}
    \label{fig:5Interface}
\end{figure*}

As shown in Fig.~\ref{fig:5Interface}, ParaTutor implements this role differentiation through two synchronized panels. The \textbf{Child Operation Panel} (D) presents an interactive visual representation of the word problem. Rather than showing solution procedures, the child-facing panel externalizes the problem situation by making relevant objects, quantities, and relations visible and manipulable. This allows children to inspect the structure of the problem, relate textual information to visual elements, and develop an initial understanding before moving into calculation. The \textbf{Parent Operation Panel} (C) serves a different role. It is the surface through which the system supports parents as mediators of the tutoring interaction, but it does not turn the system into a direct instructor for the child. At this role level, the key design decision is not only to provide different information to different users, but also to preserve the direction of tutoring: the system guides the parent, and the parent guides the child. The detailed organization of parent-facing guidance is further specified through phase-based and strategy-based scaffolding in the following sections.

Through this role-separated design, ParaTutor supports children’s problem understanding while maintaining parent-led tutoring. The system makes the mathematical structure of the problem more accessible to the child, but leaves explanation, questioning, and progression under the parent’s mediation.

\subsubsection{Scaffolding Across Tutoring Phases and Strategies}

ParaTutor organizes tutoring into three phases: \textbf{Understanding}, \textbf{Calculation}, and \textbf{Summarization}. For each problem, the parent begins by selecting scaffolding moves and entering the word problem. The system then provides two forms of support for the current phase: child-facing visual grounding and parent-facing scaffolding prompts. The parent uses this support to guide the interaction, while the child works through the problem with visual support and dialogue. Within each phase, parent and child interact iteratively, and the parent may query the system on demand for additional help. Moving to the next phase requires joint confirmation, which helps both users remain aligned and prevents premature advancement. This design preserves parent-led tutoring while keeping the child engaged in reasoning. Fig.~\ref{fig:2FlowChart} in Appendix~\ref{sec:Interaction} provides an overview of this flow.

This interaction is implemented through phase-specific agents whose outputs are constrained by both role and tutoring stage. In \textbf{Understanding}, agents prioritize segmentation and grounding prompts while withholding solution procedures. In \textbf{Calculation}, agents support error localization and guided questioning. In \textbf{Summarization}, agents support reflection, consolidation, and transfer prompts. Across phases, the orchestrator routes outputs to the appropriate parent- or child-facing surface and enforces answer-disclosure constraints.

\begin{figure}[htp]
    \centering
    \includegraphics[width=\columnwidth]{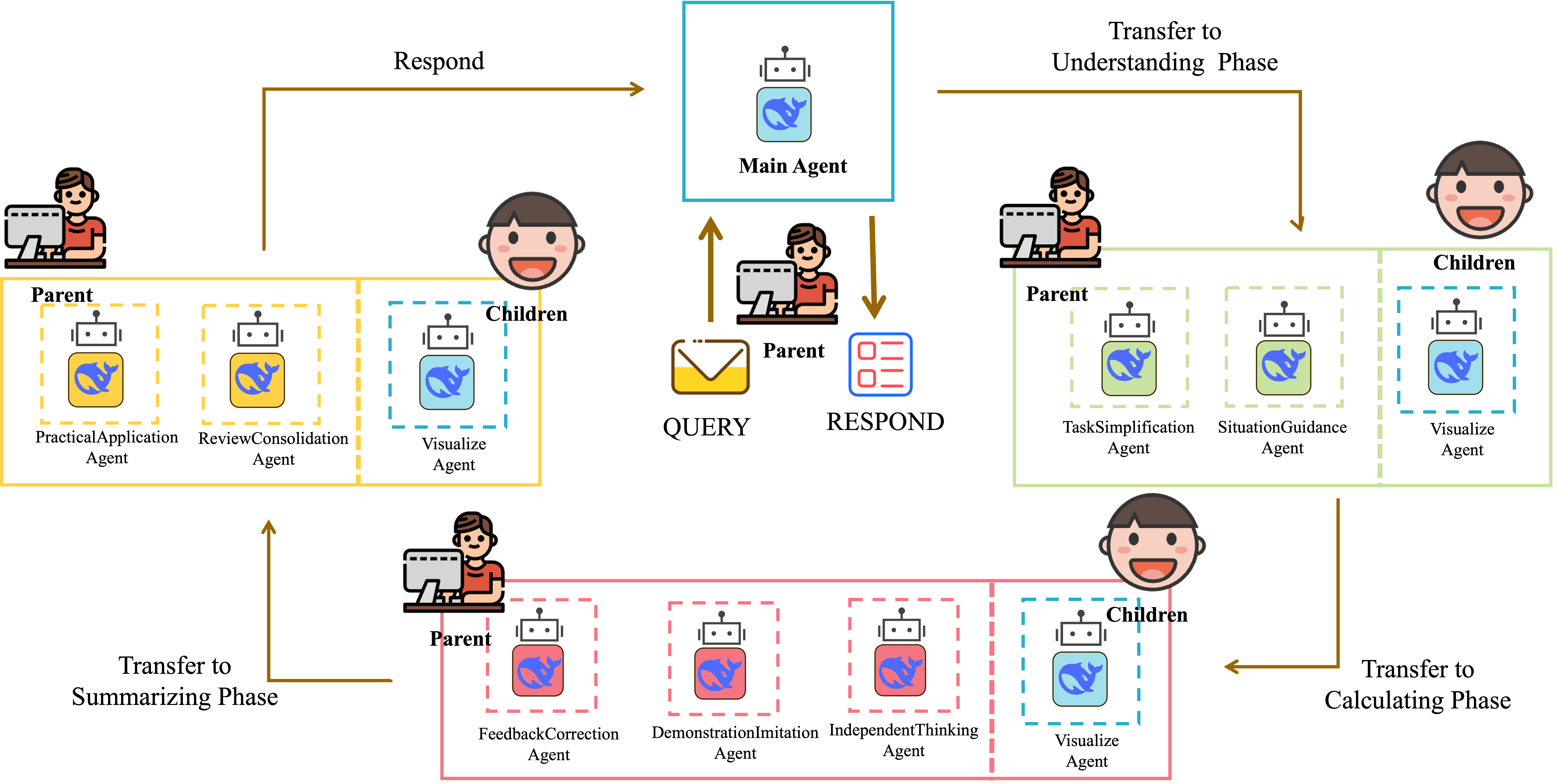}
    \caption{Implementation architecture of ParaTutor. A central orchestrator coordinates phase-specific scaffolding agents and visual grounding modules while routing outputs to role-separated interfaces. Parent-facing agents generate scaffolding prompts that support tutoring guidance, while child-facing modules provide visual grounding without revealing procedural solutions.}
    \label{fig:4_BetweenPhases}
\end{figure}

Although ParaTutor is implemented using multiple agents, the architectural novelty lies in how the system coordinates role-separated support across tutoring phases rather than in the agent decomposition itself. As shown in Fig.~\ref{fig:4_BetweenPhases}, ParaTutor consists of a \textbf{Main Agent}, a set of \textbf{phase-specific scaffolding agents}, and a \textbf{visual grounding module}. At each interaction turn, the Main Agent maintains a shared internal state that stores the current word problem, the active tutoring phase, recent interaction history, the selected tutoring mode, the child’s intermediate progress, and the current visualization state. Based on this shared state, the Main Agent invokes only the scaffolding agents relevant to the current phase, and does so sequentially rather than in parallel. For example, in the understanding phase, it prioritizes agents for task simplification and situational guidance, whereas in the calculating phase it prioritizes agents for mistake detection, feedback correction, and prompts that encourage independent thinking. Each scaffolding agent takes the updated shared state as input and returns a structured tutoring move, including both its instructional intent and its parent-facing surface realization. The Main Agent then integrates these sequential outputs into a single response shown in the parent-facing interface. 

The same shared state also drives child-facing support. When the current problem state contains entities, quantities, or relations that should be externalized, the Main Agent invokes a visual grounding module implemented as a rule-based state-to-diagram generation pipeline. Specifically, the module reads a structured problem state consisting of extracted entities, numerical attributes, and relational constraints, maps these elements to graphical primitives and layout rules, and then renders the resulting diagram as an interactive visualization. Because the visualization is generated from the evolving shared state rather than revealed as a worked solution, the child-facing interface supports interpretation and representation of the problem while withholding procedural solution steps that remain reserved for parent-facing scaffolding. Finally, phase transitions are controlled by explicit rule-based conditions over the shared internal state. Rather than allowing the LLM to decide when to move on, ParaTutor advances between the understanding, calculating, and summarizing phases only when predefined completion criteria are met, such as confirmation of problem interpretation or completion of an intermediate reasoning step.

\label{SystemArchitecture}

\section{USER STUDY}
\label{Evaluation}

The formative study characterized everyday tutoring practices and challenges, while the evaluation examines how these challenges unfolded when families used \textit{ParaTutor} in practice. To answer RQ3, we analyze how \textit{ParaTutor} used LLM scaffolding to manage breakdowns in parent-child math tutoring across three aspects. First, we examine how support was distributed across the parent, child, and LLM. Second, we examine when scaffolding entered the tutoring process, including how it became salient at breakdowns and receded across repeated problem solving. To answer RQ3, we examine how \textit{ParaTutor}'s LLM scaffolding distributed support across roles, became salient or receded during parent-child math tutoring, and fit families’ existing tutoring practices.
%metrics
We compare \textit{ParaTutor} with conventional tutoring and general LLM support to understand how role scaffolding shaped guidance delivery, child engagement, and the coordination of tutoring work.

\subsection{Experiment Design} 
Each parent-child dyad completed four tutoring \textbf{Sessions} at home. In each \textbf{Session}, the parent tutored their child using one assigned tutoring \textbf{Mode} to complete one math \textbf{Test}. To remove order effects, we used a two-level Latin Square design \cite{richardson2018use} to counterbalance tutoring \textbf{Modes} and math \textbf{Tests}. The final \textbf{Modes} and \textbf{Tests} combinations are reported in Appendix \ref{sec:AppendixE}.

We compared four tutoring modes: Conventional Tutoring (Mode A), \textit{DeepSeek} Tutoring (Mode B), \textit{ParaTutor} Aligned Tutoring (Mode C), and \textit{ParaTutor} Complementary Tutoring (Mode D). Mode A served as the no-AI baseline. Mode B provided support from a general conversational LLM that returned explanations, hints, and worked examples in response to queries from either the parent or the child, rather than delivering role-structured support. Modes C and D used \textit{ParaTutor}'s role-separated scaffolding support, with Mode C aligned to parents' existing practices and Mode D introducing less familiar, complementary strategies. This design allowed us to compare not only whether AI helps, but also how different ways of delivering LLM support reshape parent-child tutoring interaction.

The four test sets (Test 1-4) were designed by the two school teachers in the formative study to assess the same underlying mathematical logic through different real-world scenarios (\appref{sec:AppendixC}). Each test included three word problems that reflected the same three problem types, namely perimeter-based counting, two-category constraint problems, and age-related reasoning, while varying the surface context, entities, and numerical values.

\subsection{Participants and Procedure}
We recruited parent-child dyads through \textit{Xiaohongshu} and local after-school institutions between August 2024 and May 2025. Eligible participants were primary caregivers of children aged 10-12 who were actively involved in home tutoring and had access to an internet-connected computer for the remote study. The final analysis included 23 parent-child dyads, labeled E1-E23, with each family receiving 50 RMB for participation. Table~\ref{table_2} summarizes participant demographics, with parents averaging 39.1 years of age, including 6 fathers and 17 mothers. Most held a Bachelor's degree, and families varied in perceived tutoring difficulty, children's math performance, and prior familiarity with LLMs.

{\setlength{\abovecaptionskip}{4pt}
\begin{table*}[htbp]
\caption{Parental and Children's Information.}
\label{table_2}
\centering
\small
\resizebox{\textwidth}{!}{%
\begin{tabular}{|c|c|c|c|c|c|c|c|c|c|}
\hline
Label & Grade & Child Gender & Parent Gender & Parent Age & Parental Education & Tutoring Difficulty & Math Performance & Familiarity with LLMs & Province \\
\hline
E1  & 4 & Male   & Female & 38 & Bachelor       & Not difficult      & Good      & Heard and used & Guangdong \\
E2  & 4 & Female & Female & 37 & Bachelor       & Somewhat difficult & Excellent & Heard & Shandong \\
E3  & 4 & Male  & Male   & 39 & Below Bachelor & Not difficult     & Good     & Never heard & Hong Kong \\
E4  & 4 & Male   & Female & 36 & Bachelor      & Slightly easy      & Excellent & Heard and used & Sichuan \\
E5 & 4 & Male   & Female & 38 & Bachelor      & Somewhat difficult & Good      & Heard and used & Guangdong  \\
E6  & 4  & Female  & Female & 37 & Below Bachelor & Not difficult     & Good      & Heard  & Shanghai  \\
E7  & 4 & Male  & Male  & 40 & Bachelor  & Very difficult    & Below Average & Never heard & Sichuan \\
E8  & 4 & Male  & Female & 38 & Bachelor       & Slightly easy   & Good  & Heard and used & Shandong \\
E9  & 4 & Male   & Female & 37 & Below Bachelor & Somewhat difficult & Excellent & Heard & Guangdong \\
E10 & 4& Male & Female & 38 & Bachelor & Not difficult   & Good     & Heard and used & Hong Kong \\
E11 & 4 & Male   & Female & 37 & Bachelor       & Somewhat difficult & Good      & Never heard & Shanghai \\
E12 & 4 & Male   & Female & 39 & Below Bachelor & Slightly easy      & Good      & Heard & Shandong \\
E13 & 4 & Female & Female & 38 & Bachelor       & Not difficult      & Excellent & Heard and used & Sichuan \\
E14 & 5 & Male   & Male   & 42 & Bachelor       & Somewhat difficult & Good      & Heard and used & Guangdong \\
E15 & 5 & Female & Female & 41 & Bachelor       & Not difficult      & Excellent & Heard & Shanghai \\
E16 & 5 & Male   & Female & 43 & Below Bachelor & Slightly easy      & Good      & Never heard & Hong Kong \\
E17 & 5 & Female & Female & 42 & Bachelor       & Not difficult      & Good      & Heard and used & Shandong \\
E18 & 5 & Male   & Female & 41 & Bachelor       & Somewhat difficult & Average   & Heard & Sichuan \\
E19 & 6 & Female & Female & 40 & Below Bachelor & Slightly easy      & Good      & Heard and used & Hong Kong \\
E20 & 6 & Male   & Male   & 39 & Bachelor       & Not difficult      & Excellent & Heard and used & Guangdong \\
E21 & 6 & Female & Female & 38 & Bachelor     & Somewhat difficult & Good      & Never heard & Sichuan \\
E22 & 6 & Male   & Female & 40 & Below Bachelor & Slightly easy      & Good      & Heard & Shanghai \\
E23 & 6 & Male   & Female & 39 & Bachelor       & Not difficult      & Excellent & Heard and used & Shandong \\
\hline
\end{tabular}}
\end{table*}

Before the study, parents attended an online orientation session covering system use and study procedures.  Informed e-consent was obtained from parents, and assent was obtained from children. Parents and children were informed that the tutoring sessions would be video-recorded for interaction analysis. During the study, parents were asked to accompany their child throughout all sessions and prepare basic materials such as paper and pens. Each dyad completed four tutoring sessions (shown in Fig.\ref{fig:appendix-photos2}). Each session was limited to 30 minutes, with a 5-minute break between sessions. After each session, parents submitted a brief questionnaire reporting the tutoring strategy used, time spent, and photographs of the child's handwritten solutions (shown in Fig.\ref{fig:appendix-photos1}}). Upon completion of all four sessions, each parent participated in a semi-structured interview. The interview focused on (1) how the system influenced their tutoring approach to math problem-solving, (2) the effectiveness of different tutoring strategies, and (3) the overall user experience. The full interview guide is provided in \appref{sec:AppendixD}.

The analysis of the parental interview scripts followed the same procedure as in the formative study. Three researchers reviewed the entire meeting transcript to gain a general understanding. Two researchers then coded the transcript separately, categorizing the content. Any discrepancies were discussed with a senior researcher.

\subsection{Finding 1: Role-Separated Scaffolding Distributed Support Across Parents and Children}

Our first finding is that ParaTutor’s role-separated scaffolding distributed support across parents and children in complementary forms. Children received visual grounding that made the problem structure more accessible, inspectable, and discussable, while parents received text-based guidance for how to prompt, explain, and continue the tutoring interaction. This separation allowed the child to remain actively involved in reasoning while preserving the parent as the primary tutor.

\subsubsection{The Child Side}

\paragraph{Visual grounding increased children’s engagement with word problems.} For children, the emotional difficulty often appeared at the entry point of the task. Several parents described their children as becoming impatient, resistant, or ready to give up as soon as they saw a long word problem (E13, E16). 
Visual grounding changed children’s role in the tutoring interaction from primarily listening to explanations to \textbf{actively working with the problem representation}. This difference was most visible when comparing the ParaTutor modes with the two baseline conditions. In Mode A, children’s engagement depended largely on parents’ verbal explanations or self-made drawings. In Mode B, the general LLM helped parents obtain textual explanations, but children often remained peripheral while parents read, interpreted, or translated the model’s response. By contrast, Modes C and D gave children a representation they could directly inspect and manipulate.
Interaction coding reflected this shift. Using dyad-by-mode counts aggregated across the three problems in each condition, children \textbf{produced more reasoning turns} in Mode C (\textit{M} = 15.52, \textit{SD} = 3.13) and Mode D (\textit{M} = 16.83, \textit{SD} = 3.39) than in Mode A (\textit{M} = 8.61, \textit{SD} = 2.79) and Mode B (\textit{M} = 9.61, \textit{SD} = 2.80). \textbf{Child-initiated questions} showed the same pattern, increasing from Mode A (\textit{M} = 1.74, \textit{SD} = 0.92) and Mode B (\textit{M} = 2.09, \textit{SD} = 0.95) to Mode C (\textit{M} = 3.61, \textit{SD} = 1.03) and Mode D (\textit{M} = 4.09, \textit{SD} = 1.08). Children also produced more \textbf{verbal explanations}, attempts to solve, and references to problem information in the ParaTutor modes. 

Parents’ interview accounts help explain why this mattered. E13 shared, \textit{``My child used to become impatient when she saw a word problem. With the diagram, she was willing to read the sentences again and try to match them with the picture instead of giving up immediately.''} E16 similarly noted, \textit{``When he could move things on the diagram, he stayed with the problem longer. Before, he would wait for me to tell him what to do, but this time he kept trying and asked me whether his arrangement was right.''} These accounts suggest that visual grounding supported engagement not merely through interaction, but by sustaining children’s attention and encouraging them to make attempts before relying on parental explanation. However, parents also noted that visual interaction did not automatically become mathematical reasoning. E8 reflected, \textit{``At first he was very interested in moving things around, but sometimes he was just trying different positions. I still had to ask him, `What does this part mean in the problem?' Otherwise he would play with the picture without really thinking about the question.''} Thus, visual grounding created more opportunities for engagement, but productive engagement still depended on parental mediation.

\paragraph{Child-facing diagrams helped parent and child establish shared meaning for relational terms that were difficult to ground verbally.} 
The key difference was not simply whether a drawing was present, but whether relational terms that were difficult to ground verbally could be made shared through a stable and inspectable representation. In Modes A and B, parents also frequently drew temporary schematic diagrams or used gestures to explain the problem (E2, E4, E7, E11, E19). However, these parent-made drawings were often improvised during explanation and depended on parents’ ability to decide what to draw, how to map quantities onto the drawing, and how to keep the child oriented to the same reference. Several parents noted that such sketches were useful but not always sufficiently intuitive for children, especially when the problem involved sides, gaps, endpoints, corners, or counted attributes such as horns, wheels, and legs (E4, E7, E19). These terms also appeared across the evaluation problem sets, including planting problems that required children to distinguish one side from both sides, endpoints from non-endpoints, and objects from gaps, as well as animal and vehicle problems that required children to distinguish objects from counted attributes such as horns, legs, and wheels (Appendix~\ref{sec:AppendixC}).

In Modes C and D, the same terms were more often grounded directly in the child-facing diagram. When parents referred to ``both sides'' or ``each gap,'' children could point to the corresponding visual element, making it easier for the dyad to confirm what should be counted before calculation. Parents’ interview accounts illustrated this mechanism (E4, E6, E19). E19 described an interval problem, \textit{``In the beginning she kept counting the trees, but the question was really about the spaces. When the gaps were shown in the picture, I could ask her to point to each gap, and then we agreed on what should be counted.''} E4 similarly noted, \textit{``For the road problem, I was not sure whether he understood one side or both sides. With the picture, we could look at the same side and quickly decide whether the trees were planted on one side or both.''} E6 described a counted-attribute problem, \textit{``She first looked at the animals, not the horns. When she pointed to the picture, I could tell what she was counting and correct it right away.''}

\paragraph{Interactive Visual Grounding Helped Children Interpret Word Problems.}
The child-facing diagrams in Modes C and D differed from static visual explanations because ParaTutor generated interactive representations tailored to the current word problem. The interaction was especially useful for problems whose quantities were relational and difficult to imagine from text alone. In several age-problem sessions, children used the generated diagram to drag an age entity or adjust a time slider, while parents asked them to observe how the ages of two people changed together (E14, E20, E22). This made abstract relations such as ``in several years,'' ``five years ago,'' or ``twice as old'' more inspectable. In some cases, children also noticed impossible or unreasonable states during manipulation, such as a child becoming a negative age, which prompted them to reconsider their interpretation of the time relationship (E20, E22). Across the three problems in each ParaTutor mode, children interacted directly with these representations an average of 11.39 times in Mode C (\textit{SD} = 3.42) and 12.91 times in Mode D (\textit{SD} = 3.94). These interactions included dragging age markers, adjusting quantities, pointing to visual components, and using the representation to explain intermediate reasoning. The value of these actions was not merely that children touched the interface, but that they could test whether a possible interpretation was consistent with the relation described in the problem.

Parents described this interactive testing as helpful for children’s sense-making. E14 noted, \textit{``For the age problem, I first demonstrated by moving the age marker once, and then asked him to try. When he saw both ages change together, he understood that the relationship had to hold after the same number of years.''} E20 similarly described the diagram as a way for the child to test an interpretation, \textit{``He tried moving it back and forth and noticed that some positions did not make sense. That helped him understand the sentence about several years later more clearly.''} These accounts suggest that interactive visual grounding helped children interpret word problems by making hidden relational constraints manipulable and visible. However, interactivity did not always lead directly to mathematical interpretation. In some sessions, children focused on moving the slider or trying different positions without first articulating what the movement represented in the story, such as how many years had passed or why both ages changed by the same amount (E2, E5, E8, E17). Parents had to redirect them by asking what each movement meant in the problem context. This indicates that interactivity made relational constraints more visible, but still required parental guidance to turn exploration into mathematical reasoning.

\subsubsection{The Parent Side}

\paragraph{Actionable guidance helped parents recover from uncertainty and avoid repeated explanation loops.}
For parents, tutoring breakdowns in Mode A took two related forms. In some cases, parents became stuck because they did not know which method to use or could not recall how to solve the problem. In other cases, parents had an explanation but repeated it several times without the child understanding, which made the interaction increasingly tense. Session coding showed that Mode A included 11 cases in which dyads were unable to reach a justified solution within the 10-minute limit for a problem. These cases involved five families, including E2 (2 cases), E5 (2 cases), E9 (3 cases), E11 (3 cases), and E21 (1 case). Some unresolved episodes reflected parents’ method uncertainty (E2, E5, E9, E11), while others involved repeated explanations that failed to move the child forward (E5, E9, E21).

Parents described both forms of breakdown as emotionally difficult. E9 reflected, \textit{``When I could not work it out myself, I felt quite embarrassed. My child was waiting for me, and I felt that if I could not explain it, I could not really act like the parent who was supposed to guide him.''} Other parents described frustration emerging after repeated explanation. E5 noted, \textit{``I explained it several times, but he still looked confused. Then I became anxious because I did not know what else I could say.''} E21 similarly shared, \textit{``When I kept saying the same thing and she still did not understand, I could feel myself getting impatient.''} These accounts show that emotional escalation was not separate from tutoring difficulty. It emerged when parents lacked an alternative next move.

Compared with Mode A, LLM-supported modes made alternative next moves more available. In Mode B, parents could ask the general LLM for another explanation or method when they were unsure how to proceed. E2 explained that after communicating with the LLM in Mode B, \textit{``I actually learned a lot myself and helped my child understand the questions faster.''} In Modes C and D, ParaTutor further organized these next moves as parent-facing tutoring prompts, such as questions to ask, alternative explanations to try, or checks for whether the child was ready to continue. This helped parents recover from being stuck and avoid repeatedly delivering the same explanation when the child remained confused.

Overall, role-separated scaffolding redistributed the work of tutoring without replacing the parent-child interaction. Child-facing visual grounding helped children enter, inspect, and discuss the problem structure, reducing cognitive misalignment around word problems. Parent-facing guidance helped parents recover when they were stuck or repeating ineffective explanations, reducing the conditions for emotional escalation. In this sense, ParaTutor’s role separation did not simply add AI support. It placed different forms of support where they were interactionally needed, while preserving parents as guides and children as active reasoners.

\subsection{Finding 2: Parents Translated LLM Scaffolding at Potential Breakdowns into Shared Tutoring Routines}
Finding 2 shows that ParaTutor’s scaffolding worked through a parent-mediated translation process. The system surfaced phase, strategy, and language scaffolding in the parent-facing panel when the tutoring interaction risked breaking down, such as when the child seemed confused, the parent was unsure how to proceed, a method mismatch appeared, or correction could become tense. Parents then selected, rephrased, or enacted these suggestions as child-facing tutoring moves. As dyads repeated these moves across problems, scaffolding became less visible as system output and more visible as shared routines for working through word problems.

\subsubsection{ParaTutor scaffolding surfaced to prevent tutoring breakdowns.}

Qualitative coding of the parent-LLM exchanges showed that ParaTutor’s scaffolding surfaced most clearly when the tutoring interaction encountered a breakdown. Across the ParaTutor sessions, we coded 312 parent-facing scaffolding moments. These moments were not evenly distributed across ordinary tutoring talk, but concentrated around situations where parents needed help deciding how to continue. The most common forms were strategy scaffolding (104 moments, 33.3\%), language scaffolding (96 moments, 30.8\%), repair scaffolding (58 moments, 18.6\%), and phase scaffolding (54 moments, 17.3\%). 

\paragraph{Strategy scaffolding gave parents concrete ways to regulate intervention.}
Strategy scaffolding did not usually appear as abstract labels such as ``task simplification'' or ``encouraging independent thinking.'' In the parent-facing panel, it appeared as concrete suggestions for how parents could adjust their level of intervention in the child’s reasoning. In the coded parent-LLM exchanges, these suggestions ranged from more directive moves, such as breaking a problem into smaller parts or demonstrating one step, to lighter-touch moves, such as asking the child to explain their thinking before the parent intervened. For example, when parents indicated that the child was overwhelmed by a long word problem, ParaTutor often suggested a decomposition move rather than a solution. In one exchange, E12 asked what to do because the child ``did not know where to start.'' The system responded by suggesting, \textit{``Ask your child to read only the first sentence first. Then ask: what do we know from this sentence? After that, move to the next sentence and circle the number or condition it gives us.''} This type of scaffolding helped parents slow the task down and turn a dense problem statement into a sequence of smaller interpretive steps. When parents reported that the child could not connect the problem to a concrete situation, the system suggested situating the problem in a familiar context. In an exchange with E8, ParaTutor suggested, \textit{``You can first turn the problem into a story from daily life. For example, ask: if this were about lining up toys or sharing snacks, which part would represent the total and which part would represent each group?''} This did not give the child the answer, but gave the parent a way to make the mathematical relationship more imaginable.

Other strategy scaffolds helped parents decide when to intervene and when to hold back. In several exchanges, parents asked whether they should directly explain the next step after the child hesitated. ParaTutor often suggested a lighter prompt first. For example, in E7, the system suggested, \textit{``Before explaining, ask him: what have you already figured out, and what part is still uncertain? If he can name the stuck point, give only one hint rather than the full step.''} Parents of higher-performing children described this as consistent with their preference to support independence. E7 explained, \textit{``If he can already understand the concept at school, I prefer to let him try first without too much interference. I just step in when he’s stuck.''} In one exchange, E14 asked how to help the child after several failed attempts. ParaTutor suggested, \textit{``Show only the first small step on a similar example, then ask your child to do the same step on this problem. After that, ask why the two steps are similar.''} This gave the parent a demonstration move while still returning the main reasoning task to the child.

These examples show that strategy scaffolding gave parents a repertoire of intervention moves rather than a single explanation. The system helped parents decide whether to simplify, correct or hold back, depending on where the child was stuck. In this sense, strategy scaffolding supported parents in regulating the amount and form of help, rather than simply increasing the amount of help.

\paragraph{Language scaffolding helped parents turn LLM-generated methods into child-accessible explanations.}
Parents did not always struggle because the system failed to provide a method. More often, the difficulty appeared after the method was generated, when parents had to decide how to say it to the child. Several parents described this gap between “I can understand it” and “I can explain it” as a recurring problem in the ParaTutor sessions (E6, E10, E14, E18). E6 explained, \textit{``Sometimes I understood what the system meant, but if I read it directly to my child, she still would not understand. I needed it to become words that I could actually say to her.''} E18 similarly noted, \textit{``The answer was there, but the language was still too adult. I had to ask it to say it in a way a child could follow.''} Across the ParaTutor sessions, 96 language scaffolding moments were coded. These included simplifying abstract mathematical expressions into everyday wording (38 moments, 39.6\%), rephrasing a method into step-by-step child-facing language (31 moments, 32.3\%), translating an LLM-generated method into language that matched the child’s current understanding (19 moments, 19.8\%), and generating concrete examples or analogies to make a relation easier to explain (8 moments, 8.3\%). These moments show that language scaffolding did not merely provide parents with correct solutions. It helped parents convert system-generated reasoning into language that could be used in the parent-child interaction.

For example, in an age problem, E14 asked how to explain why two people’s ages must change by the same number of years. ParaTutor did not only return an equation-based method. It suggested language the parent could say to the child, such as \textit{``If one year passes, both people grow one year older. We cannot move only one person’s age. Let’s move both ages and see when the relationship becomes true.''} E14 later reflected, \textit{``I understood the method after reading it, but I did not know how to say it without making it sound like algebra. The system gave me a way to say it in ordinary language.''} A similar pattern appeared in vehicle and animal counting problems. When the LLM-generated method relied on comparison or hypothetical substitution, parents often needed help turning that reasoning into a sequence of child-facing prompts. In one vehicle problem, ParaTutor suggested that the parent begin by assuming all vehicles were motorcycles, ask the child how many wheels were missing from the total, and then explain that changing one motorcycle into a car adds two wheels. E10 described this as useful because it changed the explanation from something the parent could understand into something the child could follow: \textit{``I understood what the system meant, but if I just read it out, he would not follow. After it changed the wording, I could explain it step by step.''}

\paragraph{Repair scaffolding buffered emotionally tense correction moments.}
Repair scaffolding appeared when parents needed to respond to children’s mistakes, hesitation, or resistance without pushing the interaction into conflict. In these moments, ParaTutor suggested neutral and encouraging language that parents could adapt in their own speech. Rather than framing the child’s response as simply wrong, the system often offered sentence starters such as \textit{``You are on the right track,''} \textit{``Let’s check this part again,''} or \textit{``That is okay, let’s try another way.''} These prompts helped parents maintain a constructive tone while still directing the child back to the mathematical task.

Parents described this language as useful because emotional tension often emerged after repeated explanation or correction. E15 noted that the system’s phrasing made a visible difference during correction: \textit{``Instead of saying `wrong,' it suggested `let’s try a different way.' That made the atmosphere much better, and he was still willing to continue.''} E23 similarly described ParaTutor as creating distance in interactions that could otherwise become conflictual: \textit{``Now that my child is in adolescence, she often finds me annoying. But with the system, things feel less tense. It creates some distance so she can think without us getting into arguments.''}

This suggests that repair scaffolding did not remove the parent from the tutoring interaction. Instead, it gave parents alternative language for staying in the interaction when correction became emotionally delicate. Compared with general conversational LLM support, ParaTutor kept parents centrally involved by making guidance something they delivered and adapted, rather than something they merely retrieved from the model.

\paragraph{Repair scaffolding helped parents correct errors without escalating the interaction.} Across the ParaTutor sessions, 58 repair scaffolding moments were coded. These moments most often softened direct correction into a checking prompt (22 moments, 37.9\%), redirected the child to a specific step (15 moments, 25.9\%), offered encouragement before correction (12 moments, 20.7\%), or suggested different wording when the parent’s current explanation was not working (9 moments, 15.5\%). Rather than prompting parents to say that an answer was wrong, ParaTutor often suggested phrases such as \textit{``Let’s check this part again,''} \textit{``You are close, but one condition may still be missing,''} or \textit{``That is okay, let’s try another way.''}

Several parents described that after explaining a problem repeatedly, they became more likely to directly point out the child’s mistake, repeat the same explanation, or show impatience when the child still did not follow (E5, E9, E15, E21). E21 explained, \textit{``When I had already explained it several times and she still did not understand, I could feel myself becoming impatient. At that point, I did not know what else to say except asking her to look again.''} In these moments, the problem was no longer only mathematical. The parent also needed a way to keep the child willing to continue. Parents described repair scaffolding as useful because it gave them an alternative to repeated explanation or blunt correction. E15 noted, \textit{``Instead of saying `wrong,' it suggested `let’s try a different way.' That made the atmosphere much better, and he was still willing to continue.''} E5 similarly reflected, \textit{``When I kept explaining and he still did not understand, I would normally become anxious. The system reminded me to ask him which step he was unsure about, so I could stop repeating the whole thing.''} For E23, this also mattered in a parent-child relationship where direct correction could easily be resisted: \textit{``Now that my child is in adolescence, she often finds me annoying. But with the system, things feel less tense. It creates some distance so she can think without us getting into arguments.''}

\paragraph{Phase scaffolding helped parents regulate tutoring progression.}
54 phase scaffolding moments most often reminded parents to keep the child in problem understanding before calculation (21 moments, 38.9\%), guide calculation one step at a time (17 moments, 31.5\%), prompt the child to summarize the method after solving (10 moments, 18.5\%), or check whether the child was ready to move forward (6 moments, 11.1\%). This form of scaffolding helped parents avoid treating the solution as one continuous explanation. Instead, it made the current tutoring goal more explicit. E19 compared this with general LLM support, noting, \textit{``With DeepSeek, it just gives you the whole solution at once. But with ParaTutor, it breaks things down step by step, so that we can really work through it together.''} Phase scaffolding therefore helped parents slow down when the child had not yet understood the problem, continue step by step during calculation, and use summarization to turn a solved problem into a reusable method.

\subsubsection{Parents translated scaffolding into child-facing tutoring moves.}
Although the parent-facing panel provided prompts, explanations, repair language, and phase reminders, parents still decided whether and how to use them with their children. Interaction coding showed that across the ParaTutor sessions, 312 parent-facing scaffolding moments were observed. Parents directly adopted the suggested guidance in 84 moments (26.9\%), rephrased or adapted it in 164 moments (52.6\%), and skipped, delayed, or returned to it later in 64 moments (20.5\%). This pattern suggests that parents most often treated system guidance as material to be translated, rather than as lines to be read aloud.

This translation was visible in how parents changed the system’s wording to fit the child’s response, their own speaking style, or the immediate state of the tutoring interaction. For example, a system prompt such as \textit{``Ask the child to identify the known quantities and unknown quantity''} was often reformulated into a more conversational question, such as \textit{``What do we already know here, and what are we trying to find?''} Similarly, repair prompts were softened or personalized before being spoken to the child. E12 described this process: \textit{``I would look at what it suggested first, but I still needed to decide whether my child was ready for the next step. Sometimes I changed the wording so it sounded like how I usually talk to him.''} E20 similarly noted that the parent-facing panel was useful because it gave a direction rather than a script: \textit{``I did not read every sentence to him. I looked at the suggestion, understood what it wanted me to do, and then said it in my own way.''}

Parents also retained control over progression. Across coded phase-transition opportunities, parents moved to the next phase in 119 cases and stayed in the current phase for further explanation in 65 cases. These decisions show that scaffolding was mediated by parental judgment. Parents used ParaTutor to decide what kind of support was appropriate, but the actual tutoring move remained embedded in the parent-child interaction. In this sense, the parent-facing panel functioned less as an automated tutor and more as a backstage resource that parents translated into child-facing guidance.

\subsubsection{Translated scaffolding faded into shared routines.}

As parents translated ParaTutor’s scaffolding into their own tutoring language, some prompts gradually became part of the dyad’s routine way of working through a problem. This was most visible when the system’s strategy aligned with parents’ existing tutoring habits. In Mode C, parents often needed less negotiation to use the scaffold because the suggested moves resembled what they already tried to do, such as asking the child to restate the problem, checking one condition at a time, or summarizing the method after solving. Across sessions, aligned Mode C had a shorter average question completion time (328.02 seconds) than complementary Mode D (364.93 seconds), and this difference was significant in a paired-sample t-test ($p \approx 0.000012$). This suggests that aligned scaffolding more easily faded into the parent-child interaction because it extended familiar tutoring routines rather than requiring parents to reorganize their approach during the task.

Parents’ accounts also reflected this shift from explicit guidance to routine practice. E14 described that after using ParaTutor several times, they no longer needed to read every prompt carefully: \textit{``At first I looked at the system to know what to ask. Later I already knew that I should ask him to say what the problem means first, then check the numbers, and only then calculate.''} E7 similarly noted that the system’s phase reminders gradually became a shared sequence: \textit{``After a few problems, he also knew we would first understand the question, then calculate, then say what method we used. We did not need to argue about what to do next.''} In these cases, scaffolding faded not because support disappeared, but because the parent and child began to appropriate the structure as their own way of tutoring.

Complementary Mode D showed a different pattern. Because parents were asked to use strategies they rarely used, the scaffolds did not fade as quickly into routine. Some parents paused more often to interpret the suggested approach or decide how to explain it to the child. E12 reflected, \textit{``It was hard to change how I usually explain things. I wasn’t sure if I was helping or making it more complicated.''} This added burden helps explain why Mode D took longer on average. However, the slower coordination was not only a limitation. Several parents described complementary strategies as useful for noticing alternative ways of guiding the child. E11 explained, \textit{``It made me realize that I do not always have to explain first. Sometimes asking him to try and then discuss his thinking also works.''} Thus, aligned scaffolding supported smoother coordination, while complementary scaffolding created more friction but also expanded parents’ tutoring repertoire.

\subsection{Practical Outcomes and Usability}

Beyond the interactional findings above, we also examined whether \textit{ParaTutor} was practical for families to use during tutoring. Overall, parents reported that the interface was easy to understand and did not add substantial burden to the tutoring process. They especially noted that mode selection, parent-facing prompts, and feedback displays were clear enough to use during live parent-child interaction. E13 commented, \textit{``We found the interface intuitive, especially when selecting different modes. The seamless integration of mode selection and feedback display contributed to her learning.''} E22 similarly noted, \textit{``Once I showed my child how to use the first question, she could continue the rest independently because the prompts were so clear.''} These accounts suggest that the interface supported tutoring by keeping system guidance accessible at the moment parents and children needed it.

\begin{figure}[htp]
    \centering
    \includegraphics[width=0.75\columnwidth]{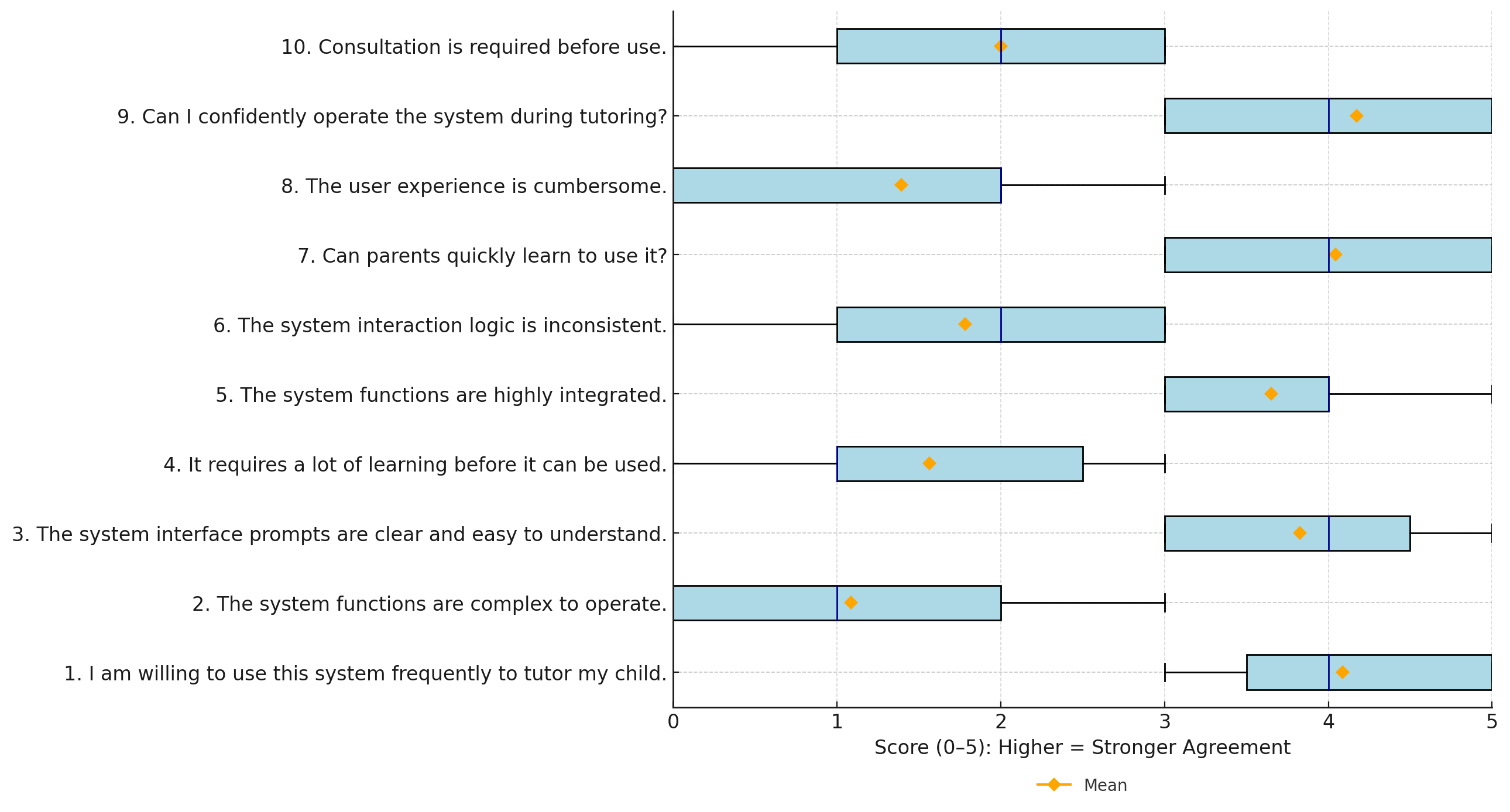}
    \caption{Perceived usability of \textit{ParaTutor} based on item-level SUS ratings.}
    \label{fig:SUS}
\end{figure}

The SUS results were consistent with this feedback. As shown in Fig.~\ref{fig:SUS}, participants gave high ratings on items related to ease of use and clarity. Item 3, which measured whether the system prompts were clear and easy to understand, and Item 1, which measured willingness to use the system frequently for tutoring, both received mean scores above 4 with relatively low variance. These results suggest that \textit{ParaTutor} was perceived as usable in the home tutoring context, where parents need to attend to both the system and their child during problem solving.

\section{DISCUSSION}
\label{Discussion}

\subsection{Scaffolding as Role Coordination in Family Tutoring}
Our findings suggest that the contribution of \textit{ParaTutor} lies not simply in providing more help, but in restructuring how help is delivered in parent-child tutoring. Prior HCI work has operationalized scaffolding through structured prompts, task decomposition, annotation, feedback, and dialog guidance across domains such as writing, programming, literature review, and healthcare communication \cite{hui2023lettersmith, ma2025dbox, palani2023relatedly, hu2024designing}. This work shows that scaffolding can help users stay oriented in complex tasks and receive context-sensitive support without being given final answers too early. However, much of this work treats scaffolding primarily as support for an individual user’s task progress or for a human-AI interaction. Parent-child tutoring introduces an additional problem because the system must support a dyad whose members have unequal but interdependent roles. Prior AI- and LLM-supported family learning systems have shown that computational support can enrich shared activities and make parent-child interaction easier to initiate. In parent-child storytelling and co-reading, for example, StoryBuddy and ContextQ support caregivers by generating story experiences or dialogic questions that parents can use to sustain conversation with children \cite{zhang2022storybuddy, dietz2024contextq}. SET-PAiREd similarly gives parents control over LLM-generated educational content by allowing them to review and revise what a learning companion robot will deliver \cite{ho2025set}. These systems are important because they avoid treating parents as irrelevant once computational agents enter the learning activity. At the same time, their primary focus is to enrich co-experience, improve conversational opportunities, or adjust parents’ level of involvement in system-supported activities.

\textit{ParaTutor} builds on this parent-involved design direction, but math tutoring introduces a more constrained form of collaborative work. The parent is not only participating in an activity with the child. The parent is expected to organize the child’s reasoning, decide when to intervene, translate methods into child-accessible language, and manage the emotional consequences of correction. This makes scaffolding a problem of role coordination rather than only activity enrichment. Two prior systems are especially close to this role-distribution problem. BrickSmart is closely related because it also supports parent-child mathematical thinking, using generative AI to help parents guide children’s spatial language learning during family block play \cite{liu2025bricksmart}. Its contribution shows that generative support can strengthen parents’ ability to guide mathematical talk in an embodied family activity. \textit{ParaTutor} extends this direction to word-problem tutoring, where the central challenge is not only eliciting spatial language but coordinating problem representation, method explanation, correction, and phase progression while the child is solving. AACessTalk is also highly relevant because it explicitly distributes support across parent and child, offering real-time guidance to parents while recommending vocabulary cards to minimally verbal autistic children \cite{choi2025aacesstalk}. \textit{ParaTutor} adopts a similar logic of role-separated support, but in a different instructional setting. Parents receive strategy, language, repair, and phase scaffolds, while children receive visual grounding that supports reasoning without being given procedural answers. This comparison clarifies the contribution of \textit{ParaTutor}. Prior family-oriented systems show that computational support can enrich shared family activities, prompt parent-child communication, and give parents more control over system-supported content. Our findings suggest that in parent-child math tutoring, LLM support also needs to coordinate the division of instructional labor. Visual grounding helped reduce cognitive misalignment when parent and child struggled to establish shared meaning for relational terms. Parent-facing language and strategy scaffolds helped parents turn LLM-generated methods into child-facing tutoring moves. Repair and phase scaffolds helped parents slow down, soften correction, and keep the interaction from becoming emotionally tense or prematurely answer-focused. This design stance treats family learning as coordinated instructional work rather than as a direct transmission of knowledge from an LLM tutor to a child.

\subsection{Preserving Parental Agency in Culturally Situated Home Tutoring}
Our findings show that parent-facing LLM support is culturally situated. In Chinese family education, parents are often deeply involved in children’s academic work, and homework tutoring can become a site where cognitive labor, emotional labor, and family authority intersect. Recent HCI work on Chinese homework interactions shows that parent-child homework sessions often involve emotional shifts, knowledge conflicts, and escalating tensions rather than only instructional support \cite{gao2025homework}. Related work on family education labor further argues that homework tutoring includes not only visible task support, but also invisible cognitive and emotional labor that parents must coordinate within the family \cite{wang2026division}. These findings help explain why, in our study, parents’ difficulty was not only that they lacked a correct solution. When parents could not explain a problem, repeated an ineffective explanation, or corrected the child too directly, their role as capable guides was also put under pressure. This pressure is especially important in a context where parental competence can be tied to instructional authority and face. Prior work on face-work shows that people actively manage how competence, responsibility, and social standing are displayed in interaction \cite{goffman1955face,hu1944chinese,mao1994beyond}. In parent-child tutoring, this face-work happens inside the family. Several parents described feeling embarrassed or losing confidence when the child was waiting for guidance but they did not know how to continue. This suggests that LLM support in home tutoring affects not only instructional accuracy, but also how parents preserve their visible role as competent caregivers and educators.

Existing HCI work on AI in Chinese family education also shows that AI systems can shift family roles when they enter educational decision-making. For example, studies of AI-supported college application decisions in China found that parents often became the primary users of AI tools, while students were sometimes less involved in interpreting or contesting AI-generated recommendations \cite{chen2024score}. This illustrates a broader CSCW concern: AI systems do not simply provide information to a family. They can redistribute who sees information first, who interprets it, and whose judgment becomes visible in the interaction. \textit{ParaTutor} addressed this issue by keeping LLM scaffolding on the parent-facing side of the tutoring interaction. Parents could consult the system, interpret its suggestions, and decide how to translate them into child-facing tutoring moves. This backstage positioning allowed parents to receive help without making the LLM the primary visible authority in front of the child. Language scaffolding helped parents reformulate LLM-generated methods into explanations they could say to the child. Repair scaffolding offered alternatives to blunt correction or repeated explanation when the interaction became tense. Phase scaffolding helped parents decide whether to slow down, continue, or summarize. In these ways, the system preserved parents’ discretion over what to say, when to intervene, and how to move the interaction forward.

For CSCW systems that introduce LLMs into family settings, the implication is that support should be designed with attention to culturally situated role relations. LLMs do not enter neutral interactional spaces. They enter relationships where authority, responsibility, care, and emotional labor are already distributed. For parent-child learning systems, making the model more capable is therefore not sufficient. Designers also need to consider where LLM support is surfaced, who is allowed to appropriate it, and how it changes the visible roles of family members during collaboration.

\subsection{Designing for Tutoring Breakdowns Beyond Correctness}
n parent-child tutoring, correctness is only one part of the work that an LLM system must support. Prior work on intelligent tutoring systems has long emphasized diagnosis, feedback, and adaptive hints as central components of tutoring support \cite{vanlehn2011relative, graesser2005autotutor}. However, CSCW and situated action research remind us that collaborative activity often depends on how people handle breakdowns, repair misunderstandings, and coordinate work in context \cite{suchman1987plans, schmidt1992taking}. In our baseline conditions, tutoring often became difficult not simply because the final answer was unknown, but because the dyad lost shared understanding, moved too quickly into calculation, repeated an ineffective explanation, or entered emotionally tense correction. These breakdowns also align with recent work on homework interactions in Chinese families, which shows that homework tutoring often involves knowledge conflicts, emotional shifts, and parent-child tension rather than only instructional support \cite{gao2025homework}.

A correctness-oriented LLM may provide a complete solution, but doing so can leave parents with the work of translating, pacing, and repairing the interaction. This matters because scaffolding is not merely the delivery of help. Classic accounts describe scaffolding as contingent support that is adjusted to the learner’s current understanding and gradually faded as responsibility shifts \cite{wood1976role}. Work on distributed scaffolding further shows that support can be spread across tools, representations, and social actors rather than located in a single tutor \cite{puntambekar2004distributed, tabak2004synergy}. ParaTutor followed this logic by surfacing different scaffolds at different breakdown points. Visual grounding supported repair of representational misalignment. Language and strategy scaffolds helped parents recover when LLM-generated methods were not yet speakable to the child. Repair scaffolds helped soften correction and return attention to a concrete step. Phase scaffolds helped parents slow down or summarize rather than moving directly toward the answer. Designing for breakdowns also changes what counts as successful LLM support. In parent-child tutoring, success is not only whether the child reaches the correct answer, but whether the dyad can sustain a productive tutoring interaction. Future LLM-supported tutoring systems should therefore monitor not only problem-solving state, but also signs of interactional trouble, such as repeated explanation loops, unresolved references, premature phase transitions, or escalating correction. Supporting these moments would move LLM tutoring from answer generation toward collaborative repair.

\section{Limitations}
\textbf{Role-preserving support still required parental labor.}
A central goal of \textit{ParaTutor} was to keep parents actively involved rather than allowing the LLM to take over tutoring. While this design helped preserve parents' guiding role, it also meant that parents still needed to interpret system guidance, decide when to intervene, translate prompts into their own language, and regulate the emotional tone of the interaction. This responsibility remained effortful for some parents. For example, E12 reflected, \textit{``I wasn’t sure if I was helping or making it more complicated,''} while E6 noted that understanding the problem did not always mean knowing how to explain the calculation. These accounts suggest that parent-facing scaffolding may be less effective for families where parents have limited time, low confidence, or limited patience for tutoring.

\textbf{The study captured short-term interactional changes rather than long-term learning outcomes.}
Our evaluation focused on short-term use of \textit{ParaTutor} as a research prototype. The study showed how role-separated and phase-gated scaffolding reshaped parent-child tutoring interactions, but it did not establish longer-term effects on children’s mathematical reasoning, strategy transfer, or sustained changes in family tutoring practices. Future longitudinal deployments are needed to examine whether these interactional benefits persist and whether repeated use changes how parents and children approach word problems over time.

\textbf{The prototype has latency and reliability constraints.}
The current prototype runs through an LLM-agent pipeline built on the \textit{DeepSeek} API, and response latency was noticeably higher than direct interaction with the base model. These delays can interrupt the pace of parent-child tutoring and weaken conversational flow. The system also depends on LLM-generated scaffolds and visual representations, which may occasionally be incomplete, inconsistent, or require parental verification. Future versions could reduce these constraints through faster orchestration, caching mechanisms, stronger quality checks, and partial offline support.

\section{CONCLUSION}

\label{Conclusion}

This study presented \textit{ParaTutor}, a multi-agent LLM-based system designed to support parent-child math word problem tutoring at home. Drawing on formative interviews with parents and teachers, we identified recurring challenges in home tutoring, including cognitive misalignment, emotional escalation, and method mismatch. \textit{ParaTutor} addresses these challenges through role-separated and phase-gated scaffolding. Children receive visual grounding that helps them inspect and reason about problem structures, while parents receive strategy, language, repair, and phase scaffolds that help them guide the tutoring interaction without being replaced by the LLM. Through our evaluation with parent-child dyads, we found that \textit{ParaTutor} redistributed tutoring support across family roles. Child-facing visual grounding increased children’s engagement and helped parent and child establish shared meaning around difficult relational terms. Parent-facing scaffolding helped parents translate LLM-generated methods into child-accessible explanations, repair emotionally tense correction moments, and regulate tutoring progression. These findings suggest that LLM support for family learning should not be designed only as direct instruction to children. Instead, LLMs can function as scaffolding infrastructure that helps families coordinate explanation, reasoning, correction, and emotional regulation during tutoring. This work contributes to HCI and CSCW by reframing LLM-supported home tutoring as a problem of role coordination in family learning. Rather than replacing parents or simply generating answers, \textit{ParaTutor} shows how LLM systems can preserve parental agency while supporting children’s mathematical reasoning. We hope this work informs future educational systems that are sensitive to the social, relational, and instructional work involved in family learning.

\section*{Acknowledgment about the Use of LLM}
The use of generative AI tools in this work is acknowledged. GPT-4o by OpenAI was used to support language refinement, including grammar and style corrections of existing manuscript text. It was also used to generate LaTeX tables from analyzed data results and to create the visual illustrations in Fig.~1 and Fig.~2 based on author-provided concepts, descriptions, and layout intentions. The DeepSeek API service was used in the implementation of the \textit{ParaTutor} system. All AI-assisted outputs were reviewed and edited by the authors. All analyses, interpretations, conclusions, and final manuscript content remain the responsibility of the authors.

The use of generative AI tools in this work is acknowledged. GPT-4o by OpenAI was used to (1) support language refinement, including grammar and style corrections of existing manuscript text, (2)convert analyzed data results into LaTeX table format, (3) create the visual illustrations in Fig.~1 and Fig.~2 based on author-provided concepts, descriptions, and layout intentions. The DeepSeek API service was used in the implementation of the \textit{ParaTutor} system. All LLM-assisted outputs were reviewed and edited by the authors. All analyses, interpretations, conclusions, and final manuscript content remain the responsibility of the authors.

\newpage
\bibliographystyle{ACM-Reference-Format}
\bibliography{01_Mathtutor}

\newpage
\appendix

\section{Interaction Flow}
\label{sec:Interaction}
\begin{figure}[htp]
    \centering
    \includegraphics[width=\columnwidth]{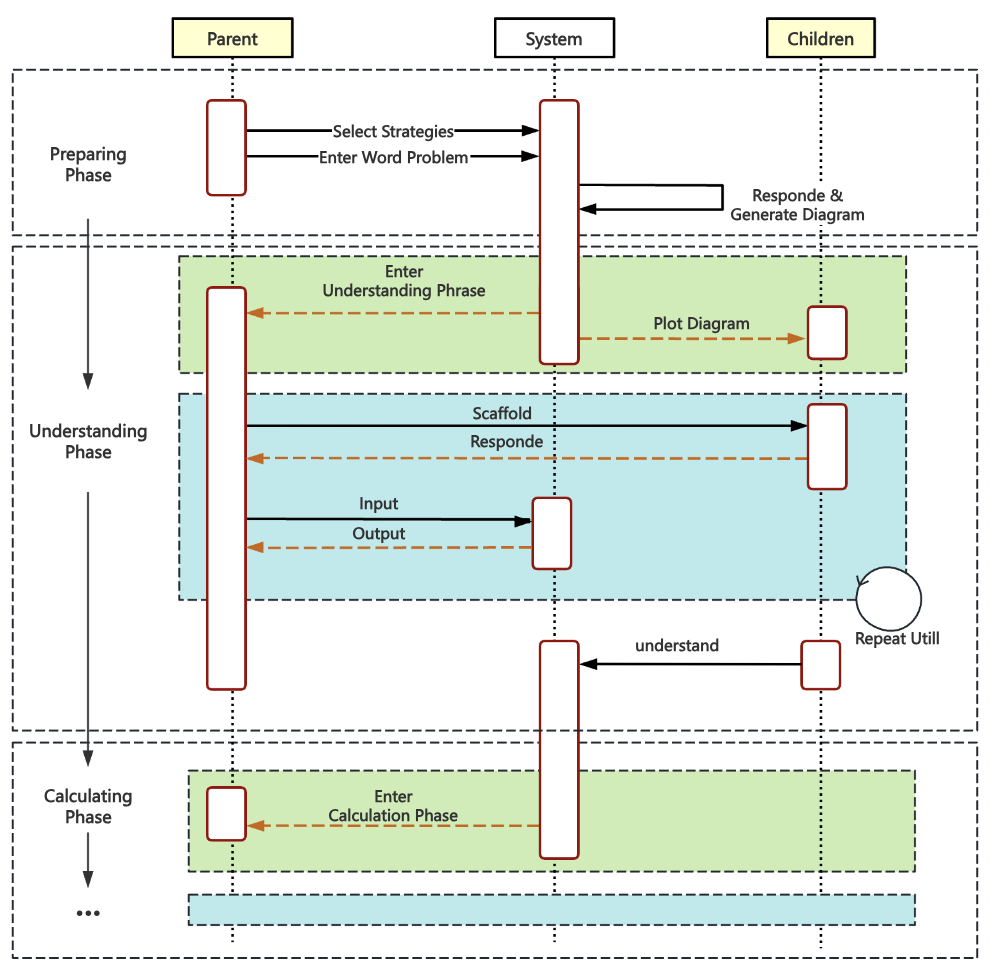}
    \caption{Parent-Child Interaction Flowchart in LLM-Assisted Math Problem Solving System.}
    \label{fig:2FlowChart}
\end{figure}

\section{Evaluation: Four Math Problem Sets}
\label{sec:AppendixC}
\textbf{Test A:}
\textbf{Practice 1}: There is a 60-meter path in the park. The plan is to plant cherry trees every 4 meters on one side of the path (not at the ends). How many saplings are needed in total? \\
\textbf{Practice 2}: There are 22 unicorns and goats in a cage, with a total of 34 horns. Unicorns have 1 horn and goats have 2 horns. How many goats are there? \\
\textbf{Practice 3}: An older sister is 12 years old and a younger sister is 4 years old. In how many years will the older sister be twice as old as the younger sister?

\textbf{Test B: }\\
\textbf{Practice 1}:  A park has a 75-meter-long main road. The plan is to plant a ginkgo tree every 5 meters on both sides of the road (at both ends). How many saplings are needed in total?\\
\textbf{Practice 2}:  Five years ago, a mother's age was seven times that of her daughter. Now, the mother's age is four times that of her daughter. How old is the daughter this year?\\
\textbf{Practice 3}:  There are 20 motorcycles and cars in the parking lot, with a total of 66 wheels. Given that motorcycles have 2 wheels and cars have 4 wheels, how many cars are there?

\textbf{Test C:}\\
\textbf{Practice 1}: A rose is planted every 6 meters along the edge of a circular flower bed with a circumference of 120 meters. How many roses are needed?\\
\textbf{Practice 2}: There are 15 ducks and 15 cows on a farm, each with 44 legs. How many ducks and cows are there?\\
\textbf{Practice 3}: A mother is 28 years older than her son, and a father is 30 years older than his son. Three years from now, the combined ages of the father and mother will be 100. How old is the son now?

\textbf{Test D:}\\
\textbf{Practice 1}:A square flower bed has 20 meters on each side. The plan is to plant a rose seedling every 5 meters (one per corner). How many seedlings are needed in total?\\
\textbf{Practice 2}:There are 18 penguins and camels, with a total of 44 legs. Given that penguins have 2 legs and camels have 4 legs, how many camels are there?\\
\textbf{Practice 3}:A father is 42 years old this year, and his son is 12 years old this year. In how many years will the father's age be 3 times his son's age minus 6?

\section{Post-study Parent Interview Outline}
\label{sec:AppendixD}
The following interview outline was designed to align with the three research questions concerning effectiveness, user experience, and feasibility. Each section contains open-ended prompts to guide qualitative data collection.\\
\textbf{1. Effectiveness of Different Tutoring Modes}
\begin{itemize}
    \item Did you observe any differences in your child’s understanding or performance across the different tutoring modes (A--D)?
    \item Which mode seemed to support learning most effectively, and why?
    \item Were there specific features (e.g., diagrams, step-by-step guidance, feedback) that helped or hindered learning in any of the modes?
    \item Compared with your usual way of tutoring, did the system help improve learning outcomes?
\end{itemize}
\textbf{2. User Experience with ParaTutor vs. DeepSeek}
\begin{itemize}
    \item How did your experience differ between using the ParaTutor system (Modes C and D) and the DeepSeek system (Mode B)?
    \item Which system provided clearer or more useful guidance to you as a parent?
    \item Did you find ParaTutor’s visual and structural feedback (e.g., strategy indicators, visual problem breakdowns) more helpful than DeepSeek’s output?
    \item Was the interaction between you and your child smoother in one system compared to the other?
\end{itemize}
\textbf{3. Feasibility and Satisfaction with Tutoring Strategies}
\begin{itemize}
    \item In ParaTutor, two modes were used: one aligning with your own strategies (Mode C), and one offering complementary strategies (Mode D). Which mode felt more feasible to implement in your home context?
    \item Did the complementary mode (Mode D) ever conflict with your preferred way of tutoring? How did you handle it?
    \item Which mode made your tutoring feel more efficient or effective?
    \item Did the system respect your role as a parent, or did it sometimes override your judgment?
    \item How satisfied were you with the overall support the system provided for your role?
\end{itemize}
\textbf{4. General Suggestions and Reflections}
\begin{itemize}
    \item What aspects of the system would you like to see improved (e.g., language clarity, pacing, visual design)?
    \item Would you prefer AI to take a leading role in tutoring, or work alongside you as a co-tutor?
    \item Would you recommend this system to other parents? Why or why not?
\end{itemize}

\section{Latin Square design}
\label{sec:AppendixE}
There are four sessions, each with a distinct combination of mode and test order. Session 1 corresponds to Mode A and Test B. After completing Session 1, participants will take a 5-minute break before moving on to Session 2, which corresponds to Mode D and Test C. Session 3 will follow with Mode C and Test D, and finally, Session 4 will correspond to Mode B and Test A. A 20-minute break is provided between each session to allow participants to rest before proceeding to the next session. Detailed data for the session configurations can be found in Table \ref{tab:mode_permutations}.
{\setlength{\abovecaptionskip}{4pt}
\begin{table}[h]
\centering
\caption{Permutations of tutoring modes (A–D) and assigned families.}
\label{tab:mode_permutations}
\small
\begin{tabular}{|c|c|c|}
\hline
\textbf{Assigned Family} & \textbf{Modes Order} & \textbf{Tests Order} \\
\hline
E1  & A-D-C-B & 2-3-4-1 \\
E2  & B-C-D-A & 4-2-1-3 \\
E3  & A-C-B-D & 3-4-1-2 \\
E4  & B-A-D-C & 1-2-4-3 \\
E5  & C-D-A-B & 4-3-1-2 \\
E6  & A-D-B-C & 3-1-4-2 \\
E7  & D-B-A-C & 2-1-3-4 \\
E8  & D-A-C-B & 1-2-3-4 \\
E9  & B-A-C-D & 2-4-3-1 \\
E10 & C-D-B-A & 3-2-1-4 \\
E11 & D-B-C-A & 3-2-4-1 \\
E12 & D-A-B-C & 1-3-2-4 \\
E13 & C-A-D-B & 4-1-2-3 \\
E14 & D-C-A-B & 3-4-2-1 \\
E15 & A-C-D-B & 2-4-1-3 \\
E16 & A-B-D-C & 4-1-3-2 \\
E17 & A-B-C-D & 3-1-4-2 \\
E18 & B-C-A-D & 1-4-2-3 \\
E19 & C-B-A-D & 2-3-4-1 \\
E20 & B-D-C-A & 4-2-1-3 \\
E21 & C-B-D-A & 3-4-1-2 \\
E22 & C-A-B-D & 3-2-1-4 \\
E23 & B-D-A-C & 4-3-2-1 \\
\hline
\end{tabular}
\end{table}

\section{Photos of Solving and Tutoring Process}
\label{sec:AppendixF}
\begin{figure}[h]
    \centering
    \includegraphics[width=0.98\linewidth]{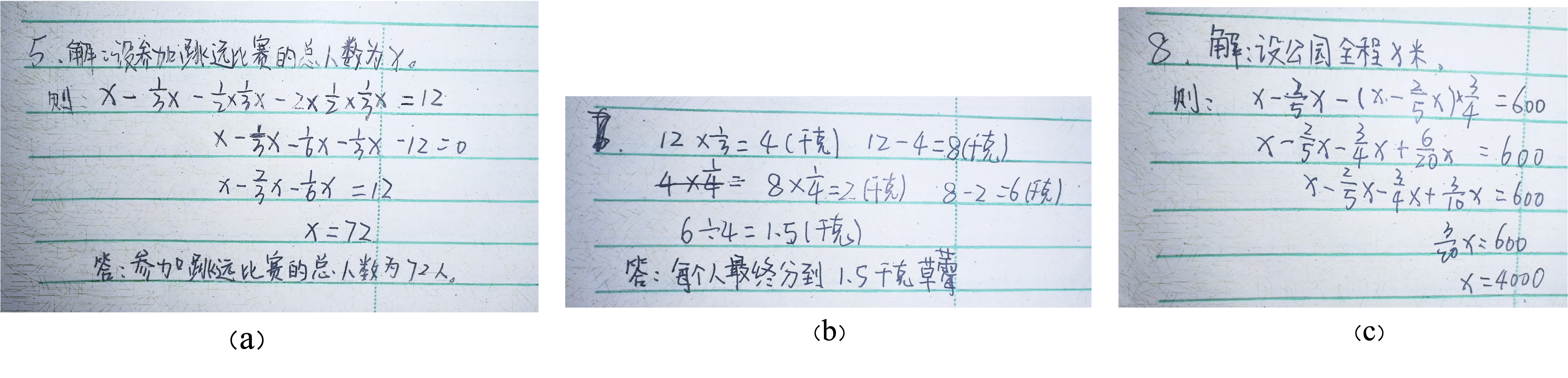}
    \caption{Participants were asked to upload three photos of the solving-process for three math problems per day on white paper in the post-questionnaire like in (a) (b) (c) shown.}
    \label{fig:appendix-photos1}
\end{figure}

\begin{figure}[h]
    \centering
    \includegraphics[width=0.5\linewidth]{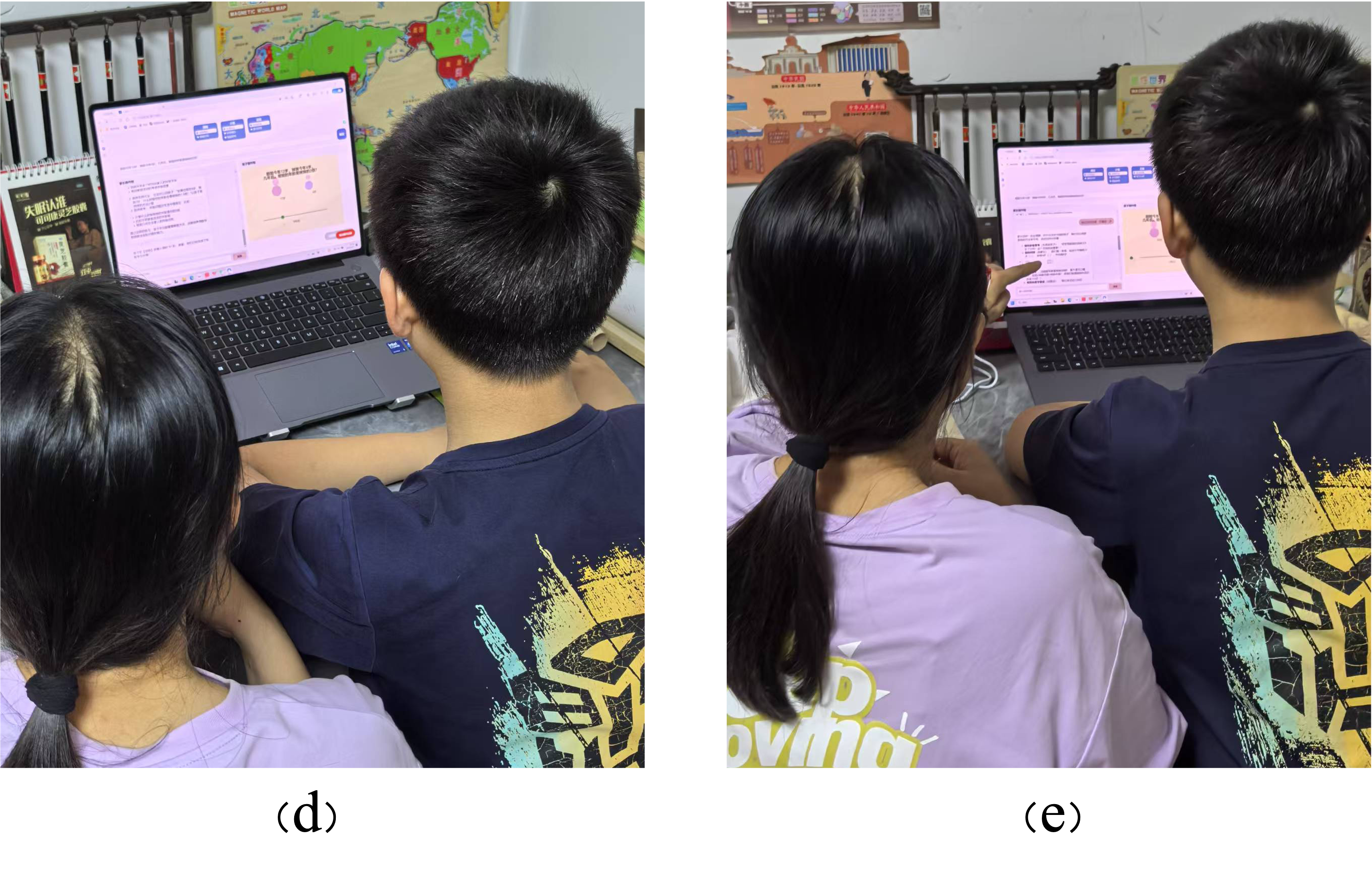}
    \caption{(d)(e) shows a participated child being faced with the e-test question we provided and is solving the question on a blank sheet of paper.}
    \label{fig:appendix-photos2}
\end{figure}

\end{document}